\documentclass[twocol]{ametsocV6.1}

\title{Estimation of AMOC transition probabilities using a machine learning based rare-event algorithm}

\authors{Valérian Jacques-Dumas,\aff{a}\correspondingauthor{Valérian Jacques-Dumas, v.s.jacques-dumas@uu.nl}
René M. van Westen,\aff{a}
Henk A. Dijkstra\aff{a,b}}

\affiliation{\aff{a}{Institute for Marine and Atmospheric research Utrecht, Department of Physics, Utrecht University, Utrecht, the Netherlands}\\
\aff{b}{Centre for Complex Systems Studies, Department of Physics, Utrecht University, Utrecht, the Netherlands}}

\abstract{The Atlantic Meridional Overturning Circulation (AMOC) is an important component of the global climate, known to be a tipping element, as it could collapse under global warming. The main objective of this study is to compute the probability that the AMOC collapses within a specified time window, using a rare-event algorithm called Trajectory-Adaptive Multilevel Splitting (TAMS). However, the efficiency and accuracy of TAMS depend on the choice of the score function. Although the definition of the optimal score function, called ``committor function" is known, it is impossible in general to compute it a priori. Here, we combine TAMS with a Next-Generation Reservoir Computing technique that estimates the committor function from the data generated by the rare-event algorithm. We test this technique in a stochastic box model of the AMOC for which two types of transition exist, the so-called F(ast)-transitions and S(low)-transitions. Results for the F-transtions compare favorably with those in the literature where a physically-informed score function was used. We show that coupling a rare-event algorithm with machine learning allows for a correct estimation of transition probabilities, transition times, and even transition paths for a wide range of model parameters. We then extend these results to the more difficult problem of S-transitions in the same model. In both cases of F-transitions and S-transitions, we also show how the Next-Generation Reservoir Computing technique can be interpreted to retrieve an analytical estimate of the committor function.}

\begin{document}

\maketitle

\section{Introduction}

Tipping elements \citep{Lenton2008, McKay2022} are subsystems of the climate system that may undergo rapid changes due to global warming. In particular, the Atlantic Meridional Overturning Circulation (AMOC) has been suggested to be in a bistable regime \citep{Stommel1961}. The AMOC collapse under freshwater forcing has since been studied in various models \citep{Rahmstorf1996} and was recently found at the higher end of the model complexity scale \citep{vanWesten2023}. Observations also suggest that the present-day AMOC is in a bistable state \citep{Bryden2011,Garzoli2013, Weijer2019, vanWesten2023}. An AMOC collapse would have a global impact, for example a decrease of the surface temperature over the Northern Hemisphere and especially Europe \citep{Jackson2015,vanWesten2024,Cini2024} and a change of rainfall patterns over tropical regions \citep{Jackson2015,Parsons2014,Chang2008}. Although doubts exist about the possibility of this collapse and the scale of its consequences \citep{Roquet2022}, the potential impact of such an event calls for an assessment of its probability of occurrence before $2100$. 

Due to its very low likelihood, this event cannot be studied by direct numerical simulations. From a theoretical point of view, Large Deviation Theory \citep{Freidlin1998} is the natural framework to study it and has recently been applied to a conceptual AMOC model \citep{Soons2023} for which the instanton, or the most likely path to the collapse, has been computed. This method also provides the optimal freshwater forcing needed to effectively collapse the AMOC. However, due partly to the required mathematical assumptions, e.g. low noise, this method is not guaranteed to be applicable to fully-coupled climate models. 

On the other hand, Trajectory-Adaptive Multilevel Splitting (TAMS) \citep{Lestang_2018} was designed to estimate the probability of any transition in fixed time. Given two sets $A$ (e.g., present-day AMOC) and $B$ (e.g., collapsed AMOC), TAMS drives any trajectory starting in $A$ towards $B$ within a chosen time. This is done with a score function that discards trajectories unlikely to transition, while cloning and resimulating promising ones. This function is key in the efficiency of TAMS: a bad score function leads to increased variance in the probability estimates and makes them more costly to compute. This is especially important in large-dimensional climate models, whose complexity makes simulated data scarce. Similar algorithms have already been successfully applied to climate problems, in particular extreme events, even in a climate model of full complexity \citep{Ragone2018}. 

The optimal score function of TAMS is the committor function \citep{Lestang_2018,Cerou2019}. It associates to any state $\mathbf{x}$ in the phase space the probability that trajectories starting from $\mathbf{x}$ reach $A$ before $B$. However, computing it exactly is already impossible in simple dynamical systems \citep{Lucente2022}. Fortunately, several data-driven methods exist to estimate this function: Analogues Markov Chain \citep{Lucente2022}, Feedforward Neural Networks, Next-Generation Reservoir Computing \citep{Gauthier2021} and Dynamical Galerkin Approximation \citep{Finkel2021}. The Feedforward Neural networks were found to give the best committor estimate \citep{Jacques-Dumas2023} but the Next-Generation Reservoir Computing (RC) technique achieved a similar performance while requiring less computational time. Other advantages of the RC include the possibility to train it online and its interpretability. 

The purpose of \citet{Jacques-Dumas2023} was to develop a benchmark for several committor estimation techniques to determine which is best fit for combination with TAMS. Here, we combine TAMS with one of them (RC) to compute transition probabilities while estimating the committor on-the-fly, using no more data than required by TAMS. We apply this technique to the same conceptual AMOC model as in \citet{Castellana}, for which there are two types of transitions: F(ast)-transitions, noise-driven transitions during which random freshwater forcings drive the AMOC strength to zero, and S(low)-transitions where the AMOC shifts to another statistical equilibrium. This model is interesting for our study because these two kinds of collapse develop on different timescales. Moreover, \citet{Castellana} already used TAMS to compute the probability of occurrence of F-transitions. These results were obtained using a prescribed score function and are compared to the TAMS-RC method. 

In section \ref{sec:methods}, we present the conceptual AMOC model, the TAMS algorithm, the committor function, and the RC method with a focus on its combination with TAMS. Then, in section \ref{sec:compare} we compare the results of the TAMS-RC method with those in \citet{Castellana}. In Section \ref{sec:s-collapse}, we extend the results of \citet{Castellana} to compute the probabilities of S-transitions. Finally, in section \ref{sec:interpret} we show how interpretable information can be extracted from the RC, and in Section \ref{sec:discussion} we summarize and discuss the results and suggest possible future improvements. 

\section{Model and methods}
\label{sec:methods}

\subsection{The Cimatoribus-Castellana model}

\begin{figure*}[t]
    \centering
    \includegraphics[width=0.75\linewidth]{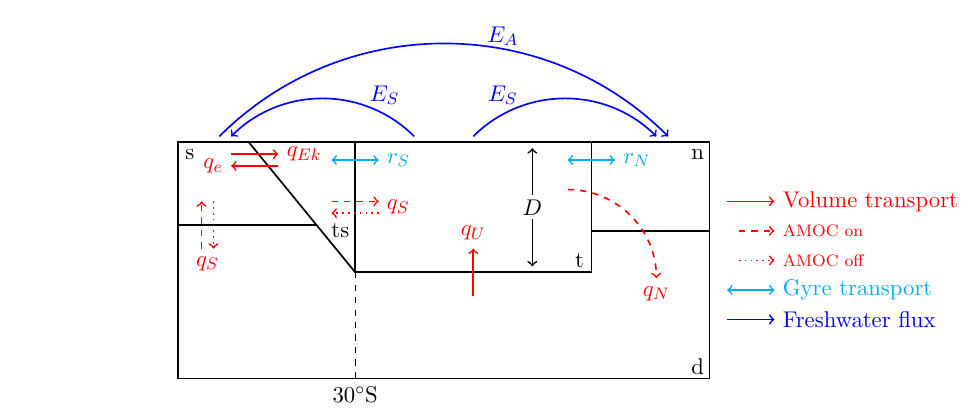}
    \caption{Summarizing picture of the AMOC model, in its version from \citet{Castellana}. Blue arrows represent the freshwater forcings, red arrows represent the volume transports and cyan arrows stand for wind-driven transports. Solid arrows are always present whatever the AMOC regime, dashed arrows correspond to the present-day AMOC regime and dotted arrows represent the fully collapsed AMOC.}
\label{fig:cimatoribus}.
\end{figure*}

Conceptual models have been very useful for understanding possible multi-stable regimes of the AMOC \citep{Dijkstra2023}. Here, we use a conceptual AMOC model introduced by \citet{Cimatoribus} with the configuration described in \citet{Castellana}. 
The Atlantic Ocean is represented by five boxes (Fig.~\ref{fig:cimatoribus}). The northern Atlantic and southern Atlantic boxes are respectively labeled $n$ and $s$. The pycnocline layer is divided into two boxes: a tropical box (labeled $t$) and a southern tropical box (labeled $ts$, located south of $30^\circ$S). Finally, there is a deep box (labeled $d$) extending below the other boxes, from the pycnocline to the ocean floor.  

The temperature of all boxes is prescribed and fixed so that the state vector $\mathbf{x}= (S_t,\ S_{ts},\ S_n,\ S_s,\ S_d,\ D)$ consists of the salinity of the boxes and the pycnocline depth $D$. The volume transports between each box are represented by two wind-driven subtropical gyres $r_s$ and $r_n$ and three quantities: $q_N$, $q_S$ and $q_U$. The quantity $q_N$ corresponds to the downwelling taking place in the Northern Atlantic, i.e. the AMOC strength. The variable $q_S$ stands for the transport in the Southern Atlantic and is the difference between the wind-driven Ekman flow ($q_{Ek}$) and the eddy-induced volume transport ($q_e$). The quantity $q_U$ represents the Ekman upwelling through the pycnocline.  

This model is forced by two freshwater fluxes. The first, $E_s$, is constant and symmetric from the box $t$ to the boxes $s$ and $n$. The second, $E_a$, is asymmetric and applied from box $s$ to box $n$. It models a physical asymmetry between both boxes that controls the strength of the AMOC while keeping the salinity budget closed. The asymmetric freshwater flux reads: $E_a(t) = \overline{E_a}(1+f_\sigma\zeta(t))$ where $\overline{E_a}$ and $f_\sigma$ are constant and $\zeta(t)$ is a white noise with zero mean and unit variance. The choice of $(\overline{E_a}, f_\sigma)$ thus determines the (stochastic) behavior of the system. The complete equations of the model and the standard configuration of the parameters are given in \citet{Jacques-Dumas2023} and not repeated here. 

In this model, the AMOC is in a bistable regime for $\overline{E_a}\in[0.06,0.35]\mathrm{\ Sv}$ ($1\mathrm{\ Sv}=10^6\mathrm{\ m}^3 $). Both its stable fixed states can be characterized in terms of volume transports. The first stable state has $q_N>q_S>q_U>0$; the downwelling in the Northern Atlantic is much stronger than the upwelling in the Southern Atlantic and through the pycnocline, corresponding to a present-day-like AMOC circulation. The second stable state has $q_N=0$ and $q_S<0$, such that the downwelling in the Northern Atlantic is shut down and the circulation in the Southern Hemisphere is reversed; this corresponds to a fully collapsed AMOC (see dotted red arrows on Fig.~\ref{fig:cimatoribus}). We call these fixed states the ``on'' state and ``off'' state of the AMOC, respectively.  A transition from the on-state to the off-state is called an S-transition and typically takes about $1,000$ years. Apart from S-transitions, F-transitions also occur in this model, during which the AMOC collapses to a state where $q_N=0$ and $q_S>0$. In that case, fast noisy freshwater inputs shut down the downwelling in the Northern Atlantic before the deep layer of the ocean can be disturbed, thus maintaining the upwelling in the Southern Atlantic. Since water accumulates above the pycnocline, this state is only transient and the AMOC either recovers or completely collapses (leading to an S-transition). The typical timescale for such a temporary shutdown is about $100$ years. 

\subsection{Trajectory-Adaptive Multilevel Sampling (TAMS)}
\label{sec:tams}

The problem of determining the probability of either F-transitions or S-transitions can be rephrased as: given two areas of the phase space $A$ and $B$, what is the probability that a trajectory starting in $A$ reaches $B$ before a time $T_{\mathrm{max}}$? For an F-transition, $A$ corresponds to a neighborhood of the AMOC-on state, $B$ to $q_N=0$ and $T_\mathrm{max}=100$ years. 

This problem is impossible to treat with direct numerical simulation, especially in the case of rare events due to the low transition probability within a limited time. However, splitting methods are designed for this kind of task. In particular, the TAMS algorithm \citep{Lestang_2018} computes the probability of occurrence of rare events at a much lower cost than Monte-Carlo simulations. This algorithm is detailed in Appendix A. At every iteration of TAMS, a score function $\Phi$ measures how close each simulated trajectory (among an ensemble) gets to $B$. The farthest ones (i.e., lowest score $\Phi$) are discarded and replaced by cloning and branching "better" trajectories. Thus, at each iteration, all trajectories get closer to $B$. TAMS is said to converge when almost all trajectories have reached $B$ (see Appendix A) and it outputs an estimate of the transition probability as well as an ensemble of trajectories, each containing a transition. TAMS thus efficiently generates an ensemble of reactive trajectories while allowing to trace back to the original transition probability of the model.

Let $\alpha$ be the transition probability from $A$ to $B$ within $T_\mathrm{max}$. Running TAMS $M>1$ times yields a distribution of its estimates $\hat{\alpha}$. Whatever score function $\Phi$ is used, $\mathbb{E}(\hat{\alpha})=\alpha$ in the limit of infinite $M$, but the variance of the distribution of $\hat{\alpha}$ depends on $\Phi$ \citep{Lestang_2018}. The optimal score function for TAMS, which minimizes this variance, is the committor function (see \citet{Lestang_2018} and \citet{Lucente2022} for its definition and properties). Appendix A also gives an expression for the optimal variance $\sigma^2_\mathrm{id}$ of $\hat{\alpha}$ when the committor is used as score function. 

\subsection{Estimating the committor function using machine learning for the AMOC model}

We aim at approaching the committor function while using only the data generated by TAMS. In \citet{Jacques-Dumas2023} we studied four data-driven methods that can perform this task. In terms of accuracy and computation time, we found Next-Generation Reservoir Computing (RC) \citep{Gauthier2021} to stand out. This method maps the states of a dynamical system onto a larger-dimensional phase space using a family of nonlinear functions. Here, we take monomials of degree $p=4$ that are products of powers of $S_n, S_{ts}, D$ and the time $t$. Our implementation of this method is presented in Appendix B and with more details in \citet{Jacques-Dumas2023}.  

Combining TAMS and RC poses the problems of initialization and the training process, since enough data is needed to obtain reliable predictions.  
The training of the RC is initialized with $1,000$ trajectories generated for $(\overline{E_a}=0.35\mathrm{\ Sv},f_\sigma=0.5)$, where the transition probability is $1$, which makes it easy to obtain transitions. For each TAMS run with parameters $(\overline{E_a}=0.35\mathrm{\ Sv},f_\sigma)$, all discarded trajectories and all trajectories reaching the target set are used to update the RC at each iteration (see Appendix B). When TAMS converges, this trained RC is used as the initial score function of TAMS for parameters $(\overline{E_a}=0.35\mathrm{\ Sv},f_\sigma-\Delta_f)$ with $\Delta_f=0.005$ and this process is repeated until $f_\sigma=0.005$. For every value of $f_\sigma$, the RC trained for $\overline{E_a}=0.35\mathrm{\ Sv}$ is used as the initial score function for TAMS with $\overline{E_a}=0.35\mathrm{\ Sv}-\Delta_{E_a}$ (with $\Delta_{E_a}=0.003\mathrm{\ Sv}$) and retrained as explained above. In this way, the RC adapts iteratively to parameters for which transitions are increasingly unlikely. This process is a form of transfer learning: information learned in an easier case (i.e., larger noise or forcing leading to a large transition probability) is transferred to a more difficult case (i.e., lower transition probability) to adapt the RC faster to this new situation. Each RC is trained using only the trajectories obtained in its own run, making them all independent. The transfer learning could also be performed by first looping over $\overline{E_A}$, then $f_\sigma$. We tested this alternative setup and found that the results hardly changed overall. 

\section{Results: F-transitions}
\label{sec:compare}

In \citet{Castellana}, the score function for TAMS is:
\begin{equation}
    \Phi^F_S(\mathbf{x}) = 1-\frac{q_N(\mathbf{x})}{q^{ON}_N},
\label{e:score}
\end{equation}
where $\mathbf{x}\in\mathbb{R}^d$ is a state in the phase space and the exponent $ON$ refers to the steady on-state of the AMOC model. $\Phi^F_S(\mathbf{x})=1$ whenever $q_N(\mathbf{x})=0$. Given a certain parameter $\rho$, we define $A$ to be a neighborhood of $q^{ON}_N$ such that $q_N(\mathbf{x})\geq q_S(\mathbf{x})$: $\ A=\{\mathbf{x}\in\mathbb{R}^n\ |\ ||\mathbf{x}-\mathbf{x}^{ON}||\leq\rho\ \&\ q_N(\mathbf{x})\geq q_S(\mathbf{x})\}$. $\Phi^F_S(\mathbf{x})$ is set to $0$ in this area. 

The score function estimated by the RC is called $\Phi_R^F$. We call TAMS-$S$ the combination of TAMS with $\Phi^F_S$ and TAMS-$R$ the combination of TAMS with $\Phi_R^F$. We sample $100$ equally spaced values of $\overline{E_a}\in[0.06,0.35]\mathrm{\ Sv}$ and $f_\sigma\in[0.005,0.5]$. TAMS uses an ensemble of $N=1,000$ trajectories, $n_c=10$ of which are discarded at each iteration; $T_\mathrm{max}=100$ years. For each couple $(\overline{E_a},f_\sigma)$, TAMS-$S$ and TAMS-$R$ are run both $M=30$ times. TAMS is stopped immediately after $\Phi^F_{\{R,S\}}=0$, before the circulation in the Southern Atlantic can be reversed, so that we are certain to sample only F-transitions with this method.

\subsection{Consistency indicator $\mathcal{C}$}

First, the transition probability estimates obtained with both methods must be consistent. We call $\alpha_S^F$ and $\alpha_R^F$ the mean probability estimates, respectively, obtained with TAMS-$S$ and TAMS-$R$. We call $\sigma_S^F$ and $\sigma_R^F$ the corresponding standard deviations. We measure the consistency of the distributions of $\alpha_S^F$ and $\alpha_R^F$ by comparing the difference of their mean values to the overlap of their $95\%$ confidence interval:
\begin{equation}
\scriptsize
\label{eq:consistency}
    \mathcal{C} = \frac{|\alpha_S^F-\alpha_R^F|}{\mathrm{min}\left(\alpha_S^F+\frac{t\sigma_S^F}{\sqrt{M}},\alpha_R^F+\frac{t\sigma_R^F}{\sqrt{M}}\right)-\mathrm{max}\left(\alpha_S^F-\frac{t\sigma_S^F}{\sqrt{M}},\alpha_R^F-\frac{t\sigma_R^F}{\sqrt{M}}\right)},
\end{equation}
where $t$ is the two-sided $95\%$ value from a Student's distribution. $\mathcal{C}<1$ ensures that both probability estimates are within the confidence interval of each other so that no method is biased \citep{Rolland2022}. 

\begin{figure*}[t]
    \centering
    \includegraphics[width=\linewidth]{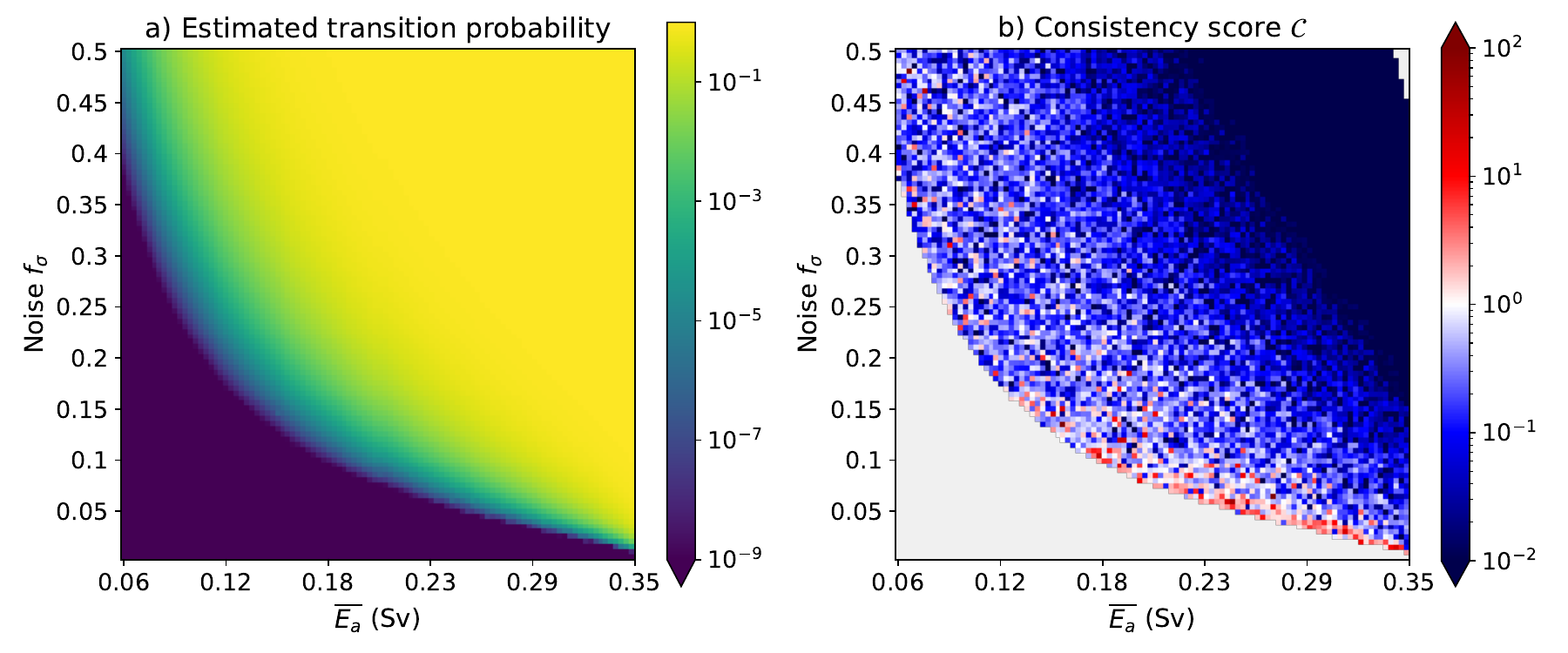}
    \caption{(a) Estimated transition probabilities using TAMS-$R$ depending on the model parameters $(\overline{E_a}, f_\sigma)$. This reproduces the main result of \citet{Castellana}, who used TAMS-$S$. \\
    (b) Consistency score $\mathcal{C}$ for every value of the couple of parameters $(\overline{E_a}, f_\sigma)$. The colorbar in this figure is centered on $1$, which means that in blue grid cells, $\alpha_S^F$ and $\alpha_R^F$ are within each other's $95\%$ confidence interval. When this is not the case, the grid cell appears in red. Gray cells mean that $\alpha_S^F$ or $\alpha_R^F$ was below the $10^{-9}$ cut-off threshold.}
    \label{fig:consistency}
\end{figure*}

Fig.~\ref{fig:consistency}b shows the consistency indicator $\mathcal{C}$. $\alpha_S^F$ and $\alpha_R^F$ are consistent for most values of $(\overline{E_a}, f_\sigma)$ since $\mathcal{C}\geq1$ only in $6.1\%$ of the cases. Fig.~\ref{fig:consistency}b can be divided into four regions. In the upper right corner, corresponding to large values of $\overline{E_a}$ and $f_\sigma$, the transition probabilities are close to $1$ and $\mathcal{C}$ is consistently below $10^{-2}$, so $\alpha_S^F$, $\alpha_R^F$ and their confidence intervals are almost identical. For lower values of $\overline{E_a}$ and $f_\sigma$, $\alpha_R^F$ and $\alpha_R^S$ are still consistent but the lighter blue shading shows that the confidence intervals do not perfectly overlap anymore. However, $|\alpha_R^F-\alpha_R^S|$ remains approximately $10$ times smaller than the overlap of their confidence intervals, so these discrepancies are not significant. Then, whatever the value of $f_\sigma$, in the region containing the smallest values of $\overline{E_a}$ such that $\alpha_R^F>10^{-9}$, all values of $|\alpha_S^F-\alpha_R^F|$ are of the same order of magnitude as the overlap of their confidence interval, although most probability estimates remain consistent. This shows that as $\alpha_R^F$ decreases, the task of the RC becomes increasingly difficult and the transfer learning propagates and amplifies biases. Finally, for $f_\sigma\leq0.13$ and the smallest values of $\overline{E_a}$ such that $\alpha_R^F\geq10^{-9}$, there is a region where inconsistencies dominate. In this regime of extremely low transition probabilities, TAMS-$R$ converges very slowly due to the biases propagated by the very long transfer learning and largely underestimates $\alpha_R^F$. It is striking to note that only $4.5\%$ of the probability estimates $\alpha_S^F\geq10^{-6}$ are inconsistent, but this number increases to $34.5\%$ for $\alpha_S^F<10^{-6}$. However, this also means that the TAMS-$R$ method is very reliable for transition probabilities down to $10^{-6}$, which is lower than often realistically needed. \citet{Collins2013} and \citet{Kriegler2009} estimate that the probability that this transition occurs within $100$ years has an upper bound of a few percent, which is many orders of magnitude above the lowest probability that our method can estimate. Moreover, this probability increases with global warming \citep{Kriegler2009,McKay2022}. Finally, studies \citep{Hall2001,Li2019,Vettoretti2022} have shown from paleoclimate simulations that such transition has a non-negligible probability, as it can occur spontaneously several times per $10^4$ years.                                                                              
\subsection{Individual variances}
\label{sec:var}

$(\sigma^2)_{\mathrm{id},\{R,S\}}^F$ (see equation (\ref{eq:id_var})) gives the optimal variance associated with the transition probability computed with TAMS-$S$ or TAMS-$R$ if it had been obtained using the committor function. So to compare the quality of TAMS-$S$ and TAMS-$R$, we compare the relative difference of their variance to their corresponding ideal variance: 
\begin{equation}
    V_{\mathrm{diff},\{R,S\}} = \frac{|(\sigma^2)_{\mathrm{id},\{R,S\}}^F-(\sigma^2)_{\{R,S\}}^F|}{(\sigma^2)_{\mathrm{id},\{R,S\}}^F}.
\end{equation}
$V_\mathrm{diff}$ is a measure of the proximity to the committor: the smaller $V_\mathrm{diff}$, the closer $\Phi_R^F$ or $\Phi_S^S$ to the committor. 

\begin{figure*}[t]
    \centering
    \includegraphics[width=\linewidth]{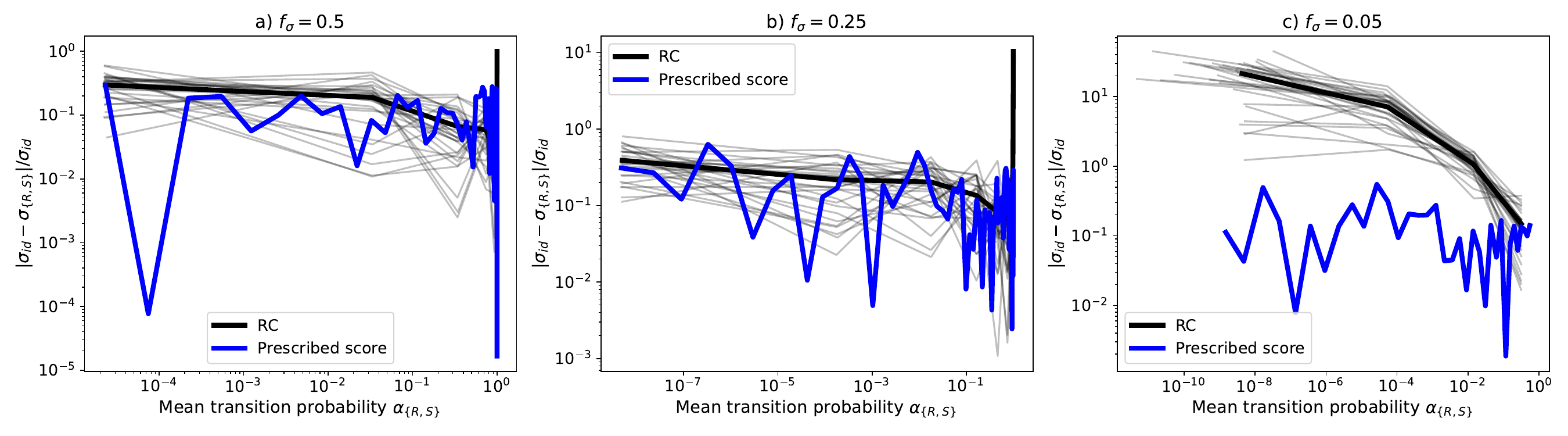}
    \caption{Each panel shows, for a fixed $f_\sigma$, the relative difference of the variance of $30$ runs of TAMS to the corresponding ideal variance $(\sigma^2)_{\mathrm{id}_{\{R,S\}}}^F$. It is plotted against the corresponding mean probability estimate $\alpha_{\{R,S\}}^F$. The blue line corresponds to $V_{\mathrm{diff},S}$ against $\alpha_S^F$. Each thin black line corresponds to $V_{\mathrm{diff},R_{i,i\in[1,30]}}$ against $\alpha_{R_{i, i\in[1,30]}}^F$. The thicker black line represents the average of the thinner black lines.}
    \label{fig:variances}
\end{figure*}

In Fig.~\ref{fig:consistency}a, TAMS-$R$ was run $30$ times using $30$ independent RCs acting as $30$ different score functions. The variance over these runs cannot be directly compared to the variance over the $30$ runs of TAMS-$S$, which uses a fixed score function. When running TAMS-$R$, the RCs are updated at each TAMS iteration, adapting them to each couple of parameters $(\overline{E_a},f_\sigma)$. These trained RCs can then be ``frozen" (their parameters are fixed and not retrained anymore) and be used as fixed score function for TAMS-$R$, so that the variance on the TAMS results when using RCs can be compared to the variance of TAMS-$S$. TAMS-$R$ is thus run $30$ times again using each of the $30$ independent RCs trained previously ($900$ runs of TAMS in total). We do this for three values of noise amplitude: $f_\sigma=0.5, 0.25$ and $0.05$. For each of these values of $f_\sigma$, results are saved for one value of $\overline{E_a}$ out of $10$ across the whole range $[0.06,0.35]\mathrm{\ Sv}$ (the computation is stopped when $\alpha_R^F\leq10^{-9}$). $V_\mathrm{diff}$ is shown in Fig.~\ref{fig:variances} for the three values of $f_\sigma$. On every panel, the blue line corresponds to $V_{\mathrm{diff},S}$ plotted against $\alpha_S^F$. Each thin black line represents $30$ runs of TAMS-$R_i$, where $i\in[1,30]$ represents each independent RC. Thus, each thin black line shows $V_{\mathrm{diff},R_i}$ plotted against $\alpha_{R_i}^F$. The thick black line is the average of the $30$ thin black lines. 

For all values of $f_\sigma$, there is a great variability between the RCs. However, for $f_\sigma=0.5$ and $f_\sigma=0.25$, all RCs are overall consistent with $\Phi^F_S$. The average of all RCs is also consistent with the curve corresponding to $\Phi^F_S$, both in terms of the mean transition probability (to be read in abscissa) and of the mean distance to the ideal score (to be read in ordinate). However, $\Phi^F_S$ shows better agreement with the committor function as $V_{\mathrm{diff},S}$ is generally lower. 

For $f_\sigma=0.05$ (Fig.~\ref{fig:variances}c), all RCs show a significant deviation from $\Phi^F_S$ and have a much greater variability between them in terms of distance to $(\sigma^2)_{\mathrm{id},R_i}^F$ than for larger $f_\sigma$. The lowest values of $\alpha_S^F$ and $\alpha_R^F$ also deviate (the left-end of the curves are not at the same point on the horizontal axis), although they remain consistent on average (thick black and blue lines) at about $10^{-8}$. This deviation illustrates the limits of transfer learning: the smaller $f_\sigma$, the longer the RC was trained by transferring information learned during easier TAMS runs. When this process takes too long, the RC may be saturated with information and has difficulty adapting to new and more complex situations. Moreover, since all RCs are independent, the longer the transfer learning process, the more likely it is to amplify biases. 

\subsection{Mean first passage times and trajectories}

Finally, the transition paths must not be biased. Two measures are considered: the mean first-passage time (MFPT) and the trajectories obtained as output of TAMS. TAMS only simulates trajectories until they reach the target set, so the MFPT can be estimated from the last time step of these trajectories.

\begin{figure*}[t]
    \centering
    \includegraphics[width=\linewidth]{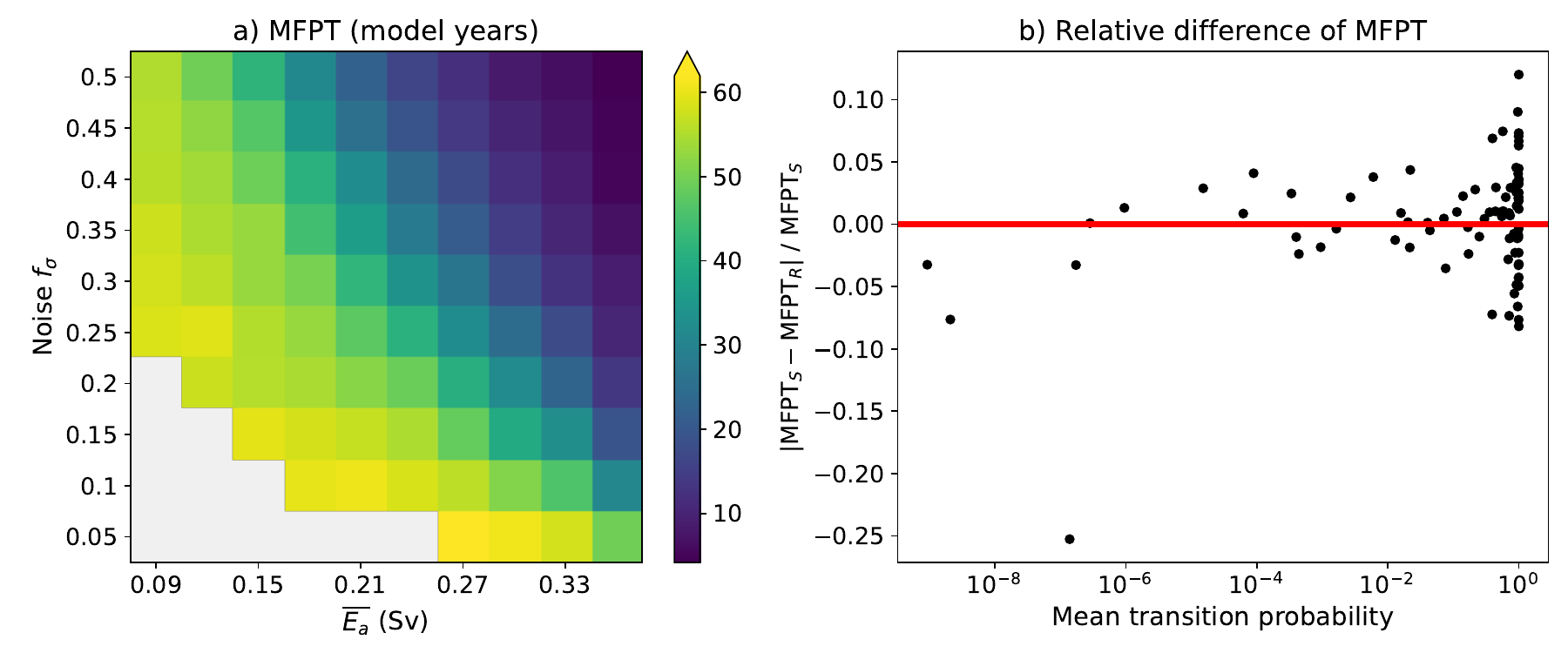}
    \caption{a): $1,000$ trajectories have been sampled from $30$ runs of TAMS-$S$. The transition times have been averaged over this ensemble for different values of $(\overline{E_a},f_\sigma)$. This panel shows the mean first-passage times $\mathrm{MFPT}_S$ thus obtained, expressed in model years.\\
    b): The MFPT has also been computed for the TAMS runs with the RC, and this panel shows the relative difference $|\mathrm{MFPT}_S-\mathrm{MFPT}_R|/\mathrm{MFPT}_S$, plotted against the mean transition probability $\alpha_S^F$.}
    \label{fig:mfpt}
\end{figure*}

Ensembles of $1,000$ trajectories were saved for several values of $(\overline{E_a},f_\sigma)$, for both TAMS-$S$ and TAMS-$R$, and the MFPTs were averaged over each ensemble. The resulting diagram corresponding to $\Phi^F_S$ is shown in Fig.~\ref{fig:mfpt}a. Fig.~\ref{fig:mfpt}b plots against $\alpha_S^F$ the relative difference of the MFPTs obtained with $\Phi_R^F$ and $\Phi^F_S$. 
Fig.~\ref{fig:mfpt}a shows that the MFPTs lie between $4.2$ model years for the largest value of $\overline{E_a}$ and $f_\sigma$ to $62$ model years for $\overline{E_a}=0.26\mathrm{\ Sv}$ and $f_\sigma=0.05$. This distribution of MFPTs is as expected, with the shortest MFPTs corresponding to the largest transition probabilities and long MFPTs associated with low transition probabilities. 

Fig.~\ref{fig:mfpt}b shows that the MFPTs predicted by $\Phi^F_S$ and $\Phi_R^F$ are in agreement. For $\alpha_S^F$ of the order of $1$, their relative error can reach $12\%$ (in absolute value). But in that case, transition times are typically shorter than $10$ years, so such an error can simply be due to the stochastic generation of trajectories. For values of $\alpha_S^F$ down to $10^{-6}$, the relative error in the estimated transition times is lower than $5\%$, therefore both methods are very consistent. The only outlier is the $25\%$ relative error on the longest transition time, which corresponds to an overestimation by $15$ years when using $\Phi_R^F$. In general, the use of $\Phi_R^F$ systematically overestimates the transition times (by $3$ to $10\%$) for $\alpha_S^F<10^{-6}$. 

\begin{figure*}[t]
   \centering
   \includegraphics[width=\linewidth]{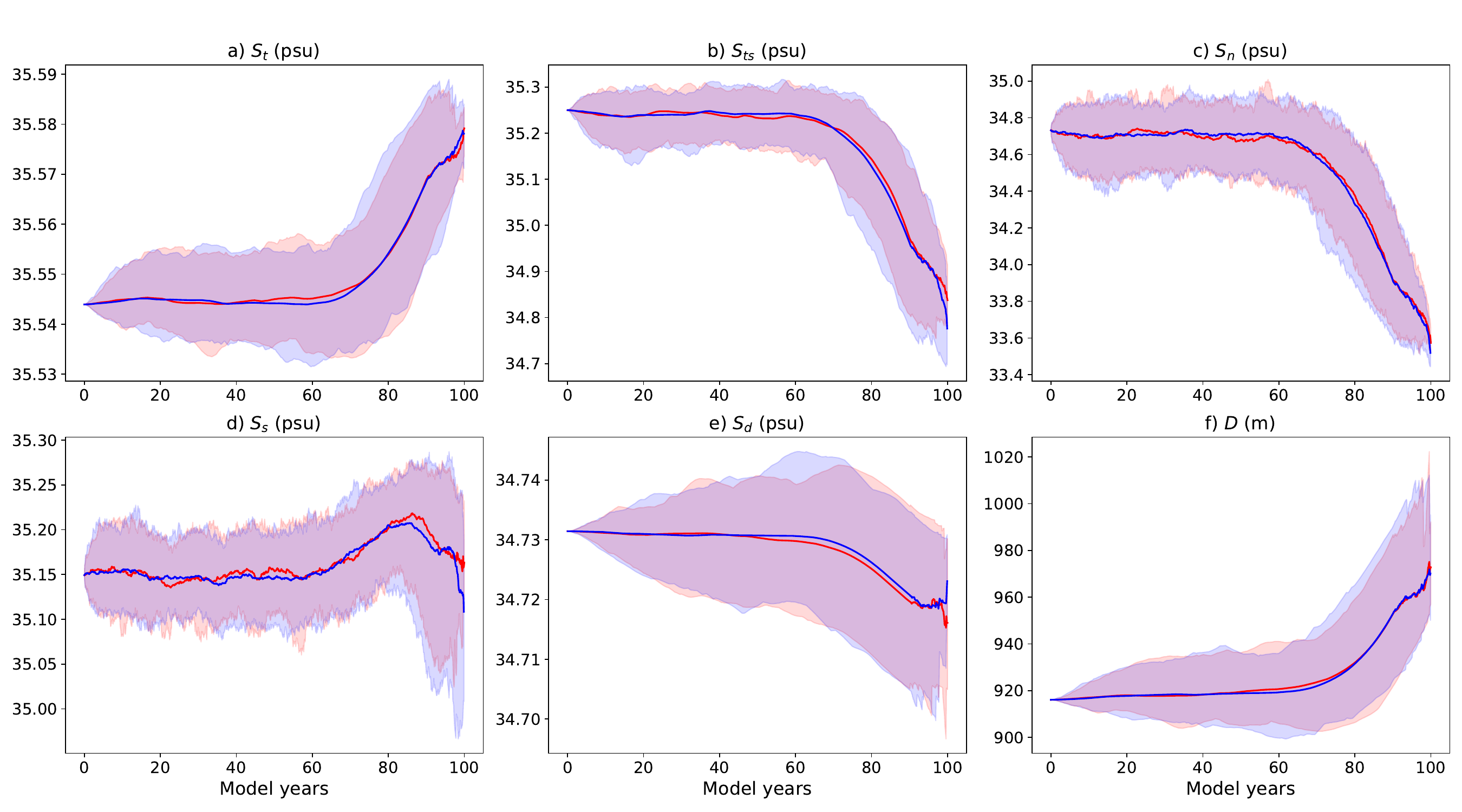}
   \caption{Ensemble of $1000$ trajectories obtained with TAMS-$S$ (in blue) and TAMS-$R$ (in red), for $(\overline{E_a},f_\sigma)=(0.288\mathrm{\ Sv},0.05)$ (corresponding to $\alpha_S^F=4\times10^{-4}$). We present here the average of the trajectories that reached $B$ between model years $90$ and $100$. The shaded areas represent the difference between the $5^\mathrm{th}$ and $95^\mathrm{th}$ percentiles in each case.}
   \label{fig:trajectories}
\end{figure*}

Then, we compare the trajectory ensembles obtained for $(\overline{E_a},f_\sigma)=(0.288\mathrm{\ Sv},0.05)$ when sampling $1,000$ trajectories from $30$ runs of TAMS-$S$ and $30$ runs of TAMS-$R$. For $f_\sigma=0.5$ and $f_\sigma=0.25$ (with similar values of $\overline{E_a}$), the results are similar as discussed below (not shown). In this case, $\alpha_S^F=4\times10^{-4}$. We have chosen these system parameters as a "worst-case scenario" because the TAMS problem is particularly difficult and the RC is more likely to be imprecise due to the repeated transfer learning processes. Moreover, due to the large variance in the MFPTs of the trajectories in this ensemble, we only consider here the trajectories whose transition time lies between $90$ and $100$ model years. This corresponds to $124$ trajectories for TAMS-$S$ and $145$ for TAMS-$R$ because, in TAMS, trajectories are left free to collapse at any time during the $100$ simulated years. We restricted ourselves to this decade to show typical collapses from trajectories within a full century. The average and the interval between the $5^\mathrm{th}$ and $95^\mathrm{th}$ percentiles of each variable are shown in Fig.~\ref{fig:trajectories}. 
The average agreement between the trajectories generated using the score $\Phi_S^F$ and $\Phi_R^F$ is very good. For all variables in the model, their averages and the distribution of $90\%$ of their values (throughout the ensemble) closely overlap. These transitions take $90$ to $100$ model years to occur, but we can see on the trajectories that the system stays around its steady on-state for the first $60$ model years. 

\section{Results: S-transitions}
\label{sec:s-collapse}

The S-transition case corresponds to a more complex problem for TAMS because the time scale of such a transition is much longer (about $1,000$ model years) and no score function can be derived from physical intuition. \citet{Castellana} also mentions this case but did not study it for a wide range of model parameters. 
For the S-transitions, we consider $10$ equally spaced values of $\overline{E_a}\in[0.09,0.345]\mathrm{\ Sv}$ (spanning the bistability regime) and $f_\sigma\in[0.05,0.5]$.
Here, as the equilibria are known, the prescribed score function \citep{Baars2021,Castellana} is:
\begin{equation}
\begin{aligned}
    \Phi^S_S(\mathbf{x}) = \eta - \eta&\exp\left(-4\frac{||\mathbf{x}-\mathbf{x}^{ON}||^2}{||\mathbf{x}^{OFF}-\mathbf{x}^{ON}||^2}\right) \\ + (1-\eta)&\exp\left(-4\frac{||\mathbf{x}-\mathbf{x}^{OFF}||^2}{||\mathbf{x}^{OFF}-\mathbf{x}^{ON}||^2}\right)
\end{aligned}
\end{equation}
where $||\cdot||$ stands for the Euclidean 2-norm and $\eta=||\mathbf{x}^U-\mathbf{x}^{ON}||/||\mathbf{x}^{OFF}-\mathbf{x}^{ON}||$ corresponds to the distance between the on-state $\mathbf{x}^{ON}$ and the unstable steady state (saddle-node) $\mathbf{x}^U$, normalized by the distance between $\mathbf{x}^{ON}$ and the steady off state $\mathbf{x}^{OFF}$. This score function makes sense from a mathematical point of view (the transition between two steady states should go through a saddle-node) but does not allow any physical interpretation. Moreover, it may be strongly biased when $f_\sigma$ is large or when the system has multiple unstable steady states. In this case, it is particularly interesting to estimate the committor function with the RC. We keep the same parameters for TAMS, except for $T_\mathrm{max}=1,000$ years, and still call TAMS-$S$ the combination of TAMS with $\Phi_S^S$ and TAMS-$R$ the combination of TAMS with $\Phi_R^S$. 

TAMS-$R$ is initialized using an ensemble of $1,000$ pre-generated trajectories with $(\overline{E_a}=0.345\mathrm{\ Sv},f_\sigma=0.5)$. But in this case, $\alpha_S^S=0.078$ so to prevent any bias, $500$ trajectories don't contain a transition (obtained by direct numerical simulation) and $500$ trajectories contain one (obtained with TAMS-$S$). Otherwise, due to the rarity of transitions, $1000$ trajectories would contain too few of them and such unbalanced training dataset would hinder performance.

\begin{figure*}[t]
    \centering
    \includegraphics[width=\linewidth]{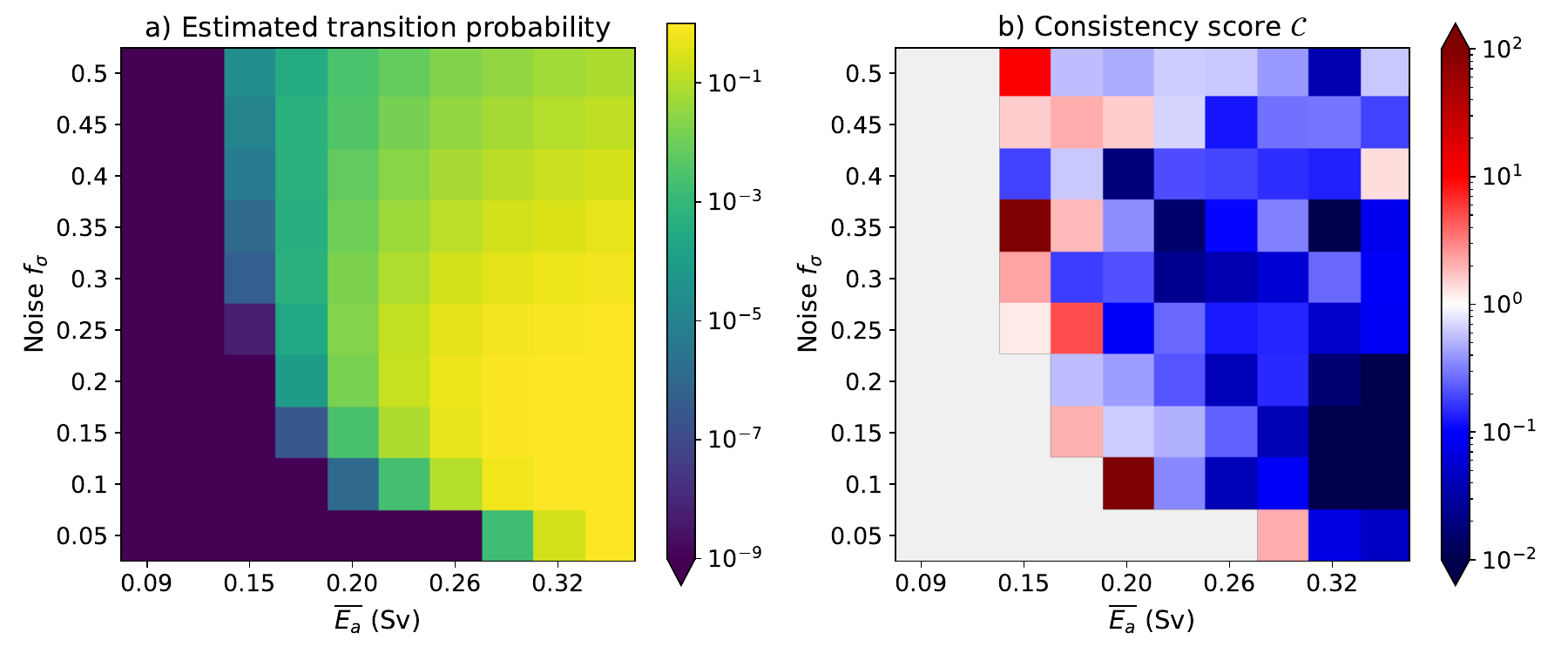}
    \caption{a) Mean estimated probability to transition from $\mathbf{x}^{ON}$ to $\mathbf{x}^{OFF}$ within $1000$ years, over the whole range of values of $(\overline{E_a},f_\sigma)$. This probability was averaged over $30$ runs of TAMS-$R$. \\
    b) Consistency score $\mathcal{C}$ for every value of the couple of parameters $(\overline{E_a}, f_\sigma)$, calculated for the transition from $\mathbf{x}^{ON}$ to $\mathbf{x}^{OFF}$. The colorbar in this figure is centered on $1$, which means that, in blue grid cells, $\alpha_S^S$ and $\alpha_R^S$ are within each other's $95\%$ confidence interval. When this is not the case, the grid cell appears in red. Gray cells mean that $\alpha_S^S$ or $\alpha_R^S$ is below the $10^{-9}$ cut-off threshold.}
    \label{fig:S_diagrams}
\end{figure*}

Figure~\ref{fig:S_diagrams}a shows the diagram of the mean transition probabilities from the AMOC on state to the AMOC off state, computed with $\Phi_R^S$. Fig.~\ref{fig:S_diagrams}b presents the consistency score $\mathcal{C}$ between TAMS-$S$ and TAMS-$R$.
Figure~\ref{fig:S_diagrams}a shows that, as expected, the smaller the freshwater forcing $\overline{E_a}$, the smaller the transition probabilities, as for the F-transitions. However, now, a large noise amplitude $f_\sigma$ tends to prevent the transition to $\mathbf{x}^{OFF}$. This can be understood by considering the basin of attraction of $\mathbf{x}^{ON}$. Since this basin is very narrow, it only requires a small noise to exit it and be attracted to $\mathbf{x}^{OFF}$. In this setting, a small noise already allows the system to switch basins of attraction and does not disturb it from its natural deterministic drift toward $\mathbf{x}^{OFF}$. On the other hand, a large noise strongly disturbs the system and may force it in directions that do not favor a transition to $\mathbf{x}^{OFF}$, thus lowering the probability of reaching it within $1,000$ years. The cut-off to compute transition probabilities with TAMS was here as well set to $10^{-9}$ but TAMS-$R$ reaches such a low probability only for $(\overline{E_a}=0.15\ \mathrm{Sv},f_\sigma=0.25)$. In all other cases, it fails to compute transition probabilities lower than $10^{-6}$. This failure of the RC for very low transition probabilities may be explained by a less efficient transfer learning process than in the case of the transient collapse. Indeed, the gap between successive sampled parameters $(\overline{E_a}, f_\sigma)$ is $10$ times larger than in the previous case, making it more difficult for the RC to adapt to new dynamics.

However, this also means that even under these poor learning conditions, the RC can still estimate transition probabilities down to $10^{-6}$ for a wide range of parameters $(\overline{E_a}, f_\sigma)$. Fig.~\ref{fig:S_diagrams}b demonstrates that the predicted probability $\alpha_R^S$ is consistent with $\alpha_S^S$ (so $\mathcal{C}<1$, shown in blue) for $83\%$ of the parameter couples. In all cases where this is not verified (except for one), the inconsistently estimated transition probabilities are the lowest for a given value of $f_\sigma$.

\begin{figure*}[t]
    \centering
    \includegraphics[width=\linewidth]{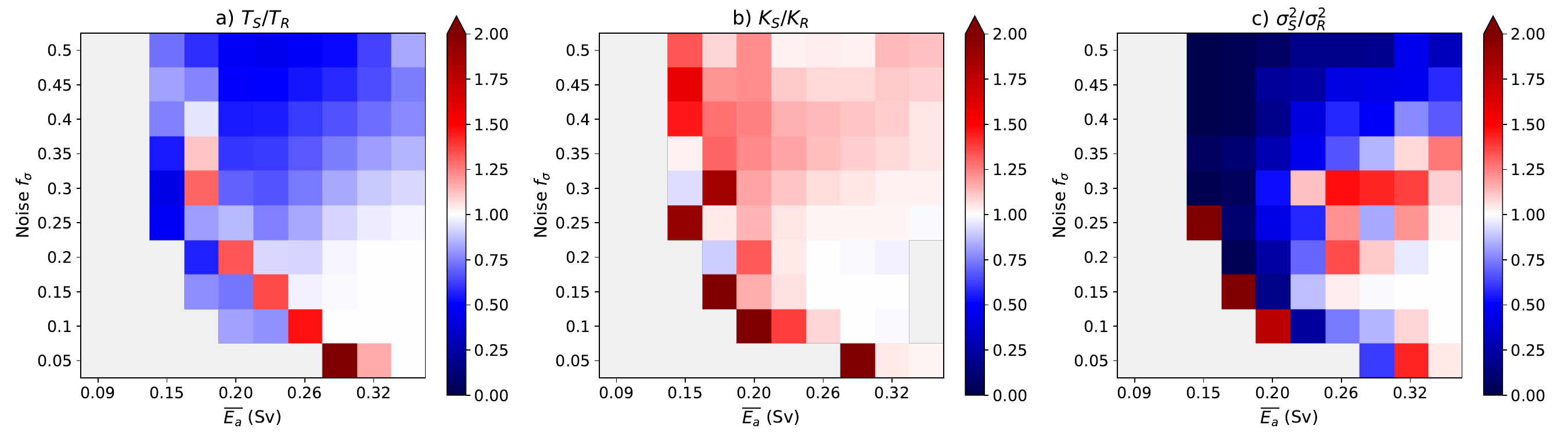}
    \caption{(a) Ratio of the number of timesteps computed by TAMS-$S$ compared to TAMS-$R$. Red values mean that TAMS-$R$ require fewer timesteps, blue values mean the opposite. In all panels, the gray color indicates that $\alpha_S^S$ or $\alpha_R^S$ is below the $10^{-9}$ cut-off threshold.\\
    (b) Ratio of the number of iterations of TAMS-$S$ compared to TAMS-$R$. Red values mean that TAMS-$R$ requires fewer iterations, blue values mean the opposite. \\
    (c) Ratio of the variance in probability estimates obtained with TAMS-$S$ compared to TAMS-$R$. Red values mean that TAMS-$S$ produces a larger variance, blue values mean the opposite. }
    \label{fig:indicators_S}
\end{figure*}

Figure~\ref{fig:indicators_S}a compares how many time steps are computed by both methods. Since each iteration of TAMS consists in integrating trajectories, i.e. simulating many time steps, we summed them up during the whole TAMS process to obtain an indicator of its computational cost. For high probabilities, both methods are, as expected, equivalent because TAMS converges very fast. When they are not equivalent, TAMS-$R$ almost always leads to the computation of more time steps, which is shown in blue in Fig.~\ref{fig:indicators_S}a. The difference is even quite large: in $50\%$ of the cases, TAMS-$R$ leads to the computation of more than $40\%$ more time steps compared to TAMS-$S$. 

However, the smaller number of time steps computed by TAMS-$S$ hides the fact that the latter takes more iterations to converge than TAMS-$R$. The number of TAMS iterations for both methods is presented in Fig.~\ref{fig:indicators_S}b, and for $90\%$ of the parameters, the number of iterations of TAMS-$R$ is smaller than that of TAMS-$S$. Even when it is larger, the difference is always less than $3\%$ of the number of iterations of TAMS-$R$ (except in two cases where this difference reaches up to $10\%$ of the number of iterations of TAMS-$R$, for probabilities lower than $10^{-4}$). For half of the parameter values, this difference corresponds to more than $15$ more iterations performed by TAMS-$S$ but can reach more than $100$ more iterations in $20\%$ of the parameter values. The fewer iterations of TAMS-$R$ are consistent with its larger number of computed time steps: to converge in fewer iterations, trajectories must be branched earlier (within $[0,T_\mathrm{max}]$) so their resimulation involves the computation of more time steps. More importantly, this also suggests a better score function, since detection of promising and wrong trajectories occurs earlier within the time interval $[0,T_\mathrm{max}]$. 

Figure~\ref{fig:indicators_S}c represents the variance around the probability estimates for both methods. As in the other panels, both methods are similar for large transition probabilities: the difference in the values of $(\sigma^2)_R^S$ and $(\sigma^2)_S^S$ is less than $5\%$ of $(\sigma^2)_S^S$ only for transition probabilities larger than $0.5$. $(\sigma^2)_R^S$ is only smaller than $(\sigma^2)_S^S$ for transition probabilities $\alpha_R^S\geq0.08$. Otherwise, TAMS-$S$ systematically yields a smaller variance, with a difference that can be quite large: $\alpha_R^S$ is more than $50\%$ larger than $\alpha_S^S$ for half of the parameter values. As is also the case in the other panels of Fig.~\ref{fig:indicators_S}, the smaller the transition probability, the larger this discrepancy and, here, the larger $(\sigma^2)_R^S$ compared to $(\sigma^2)_S^S$. As we move towards the left of the diagram, the difficulty of the transfer learning does not allow the RC to adapt to new dynamics, and its estimates are increasingly scattered. For even lower probabilities, TAMS-$R$ cannot even converge in many of the $30$ runs but gives a correct probability estimate in others, yielding a very large variance. This also explains why, to the very left of the diagram, there are some cases where $(\sigma^2)_R^S<(\sigma^2)_S^S$: in this case, there are so many TAMS runs that did not converge and returned a probability of $0$ that the prediction of TAMS-$R$ becomes quite consistent. Note that the variances presented in Fig.~\ref{fig:variances} were computed by running TAMS $30$ times for each of the $30$ independently trained RCs whereas the variance plotted in Fig.~\ref{fig:indicators_S}c is directly computed from $30$ runs of TAMS, without freezing the RCs. 

\section{Interpretability of the RC}
\label{sec:interpret}

\begin{figure*}[t]
    \centering
    \includegraphics[width=\linewidth]{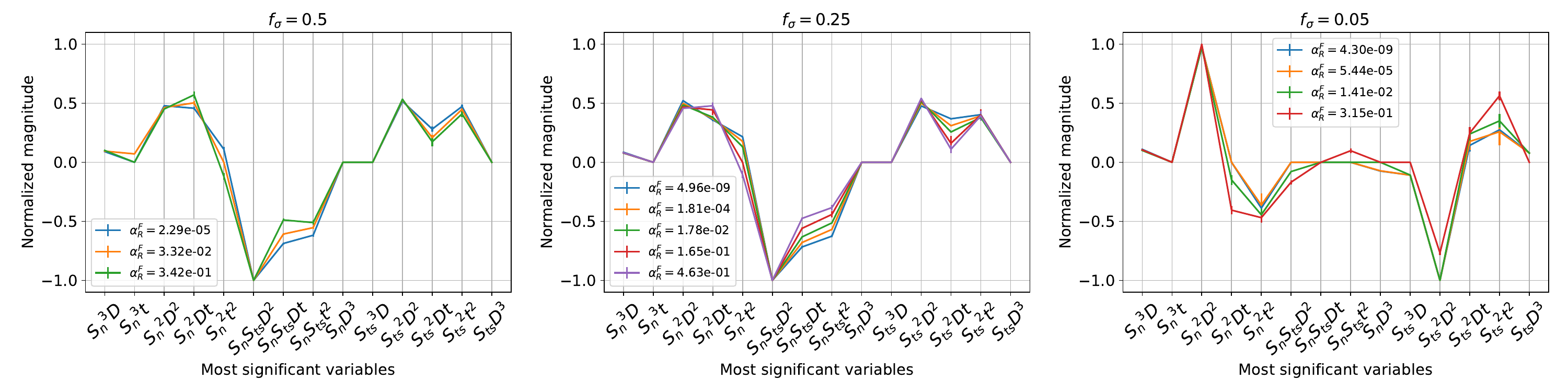}
    \caption{Each panel shows, for a fixed $f_\sigma$, the terms with largest magnitude in the matrix $W_\mathrm{out}$ of the RCs. From the $30$ RCs trained in Sect.3b, we extract the $10$ terms of $W_\mathrm{out}$ with largest magnitude. Their magnitude is then normalized to that of the largest term and we only keep the terms that are most significant for more than $20$ out of $30$ RCs. Each curve corresponds to a value of the parameter $\overline{E_a}$ (the legend gives the corresponding transition probability computed with TAMS-$R$) and they show the average magnitude of each term across the $30$ RCs and the errorbars are the corresponding standard deviations.}
    \label{fig:interpretation}
\end{figure*}

An important advantage of RCs is that we can read which term of their output matrix $W_\mathrm{out}$ contributed most to the committor estimation. In the case of F-transitions, we consider the $30$ runs of TAMS-$R$ performed at $f_\sigma=0.5, 0.25$ and $0.05$. For each of the $30$ RCs, we select the $10$ largest terms of $W_\mathrm{out}$ in absolute value and normalize them, so that the largest coefficient in each RC has a magnitude of $1$. Then, we only keep the terms that are common in more than $20$ RCs (the results are robust when varying this number between $15$ and $25$), to compare how the most significant terms are ranked by each RC. Figure~\ref{fig:interpretation} presents these relative magnitudes with their standard deviation across the $30$ RCs. Each curve corresponds to a value of $\overline{E_a}$ (or equivalently to a grid cell in Fig.~\ref{fig:consistency}a). For a given value of $f_\sigma$, they were sampled in the following way: starting from the smallest value of $\overline{E_a}$ such that $\alpha_R^F\geq10^{-9}$, we retained one value of $\overline{E_a}$ out of $10$ such that $\alpha_R^F\leq0.5$. By imposing a certain interval on the transition probabilities, we make sure that all the presented curves roughly correspond to similar dynamics of the model despite the different values of $(\overline{E_a},f_\sigma)$.  

Figure \ref{fig:interpretation} shows that, for all values of $f_\sigma$, the most important terms of all RCs are consistent: few error bars on the magnitude of every term are visible. All RCs thus extract a meaningful structure from the data and they all agree on the ranking of the most significant terms. Furthermore, for each $f_\sigma$, the structure of the matrix $W_\mathrm{out}$ converges as the transition probabilities become very small, as shown by the overlap of the curves obtained for several values of $\overline{E_a}$.

These curves also provide an analytical expression of the learned score function. For $f_\sigma=0.5$ and $f_\sigma=0.25$, $3$ interesting terms are highlighted: $S_nS_{ts}D^2$ (with a magnitude of $-1$) and $S_n^2D^2$ and $S_{ts}^2D^2$ (with a magnitude of about $0.5$). This naturally leads to a score function $\Phi_R^F\propto(S_n-S_{ts})^2D^2/2$. In both cases, we can also add the terms $(S_{ts}-S_n)S_{ts}t^2/2$ and $(S_n^2+S_{ts}^2/2-S_nS_{ts})Dt/2$ although these terms are not exactly contained in $W_\mathrm{out}$ because the coefficients of this matrix are not exactly $1$, $0.5$, or $0.25$. We still obtain a first estimate of an analytical result for $\Phi_R^F$, for $f_\sigma=0.5$ and $f_\sigma=0.25$:
\begin{equation}
\label{eq:est_comm}
\begin{aligned}
    \Phi_R^F(S_n,S_{ts},D,t) \propto\ &a_1 \frac{D^2}{2}(S_n-S_{ts})^2 \\
    - &a_2 \frac{t^2}{2}S_{ts}(S_n-S_{ts}) \\
    + &a_3 \frac{Dt}{2}\left(S_n^2+\frac{S_{ts}^2}{2}-S_nS_{ts}\right)
\end{aligned}
\end{equation}
for certain ${\cal O}(1)$ coefficients $a_i,\ i = 1,2,3$ with appropriate dimensions to make the expression consistent. 

Due to the normalization of $W_\mathrm{out}$, this expression is only a proportionality relation. Moreover, the used implementation of the RC by \citet{Gauthier2021} imposes a fixed choice of degree of the terms in the nonlinear part of the feature vector and we found $p=4$ to work best \citep{Jacques-Dumas2023}. Although $\Phi_S^F \propto q_N\propto(S_n-S_{ts})D^2$ cannot be reproduced (because $q_N$ is of degree $3$), we can compare this expression to that of $\Phi_R^F$. The RCs retrieve a similar dependence on $D^2$ and $(S_n-S_{ts})$. This salinity difference drives the density gradient between the boxes n and ts, which, in turn, controls in this model the AMOC strength $q_N$ \citep{Cimatoribus}, implying that the salt advection drives the AMOC strength and that the salt advection feedback is key in the AMOC collapse. The presence of $(S_n-S_{ts})$ in all terms of $\Phi_R^F$ shows that the RC is able to capture the importance of this mechanism. $\Phi_R^F$ also involves time, which was to be expected from the score function. Indeed, TAMS returns a transition probability conditioned on a time horizon $T_\mathrm{max}$ so the score function is ideally time-dependent. In particular, the third term of Eq.~\ref{eq:est_comm} combines $D$, $t$, and a term looking like $(S_n-S_{ts})^2$, which would not be possible if the degree of the monomials was $p<4$.

Finally, the expressions for $f_\sigma=0.5$ and $f_\sigma=0.25$ are strikingly close. This is partly due to transfer learning, but, considering that the RCs are constantly independently updated, the conservation of this structure across different values of $\overline{E_a}$ and $f_\sigma$ shows the robustness of this method. However, the different structure of $\Phi_R^F$ for $f_\sigma=0.05$ (third panel of Fig.~\ref{fig:interpretation}) shows that the RCs are still able to adapt. In this case, two monomials stand out: $S_n^2D^2$ and $-S_{ts}^2D^2$. This suggests a score function $\Phi_R^F(S_n,S_{ts},D,t)\propto(S_n-S_{ts})(S_n+S_{ts})D^2$. Other significant terms in these curves are $-(S_n^2-S_{ts}^2)Dt/4$ and $(S_n^2-S_{ts}^2/2)t^2/2$, leading to a new analytical expression of the score function obtained by the RCs:
\begin{equation}
\begin{aligned}
    \Phi_R^F(S_n,S_{ts},D,t) \propto\ &b_1 D^2(S_n^2-S_{ts}^2) \\
    - &b_2 \frac{Dt}{4}(S_n^2-S_{ts}^2) \\
    - &b_3 \frac{t^2}{2}\left(S_n^2-\frac{S_{ts}^2}{2}\right)
\end{aligned}
\end{equation}
for certain ${\cal O}(1)$ coefficients $b_i,\ i = 1,2,3$. 
The function $\Phi_R^F$ for $f_\sigma=0.05$ is different from the score function obtained for larger values of $f_\sigma$, which is not surprising as a data-driven score function is expected to depend on the model parameters. Moreover, $f_\sigma$ already appears in the equations of the committor \citep{Lucente2022}, so this function is expected to depend on it. Finally, despite the different structure of $\Phi_R^F$ for $f_\sigma=0.05$, its dependence on $(S_n-S_{ts})$ is conserved, although always multiplied by $(S_n+S_{ts})$, stressing more the importance of the salt-advection feedback. The dependence on three terms driven by $D^2$, $Dt$ and $t^2$ is also conserved.

Such an analysis may be very useful in more complex cases and larger-dimensional models to understand the possible structure of a good score function. We test this point by interpreting the output of $W_\mathrm{out}$ in the case of the AMOC S-transitions (see Sect.~\ref{sec:s-collapse}). 

\begin{figure*}[t]
    \centering
    \includegraphics[width=\linewidth]{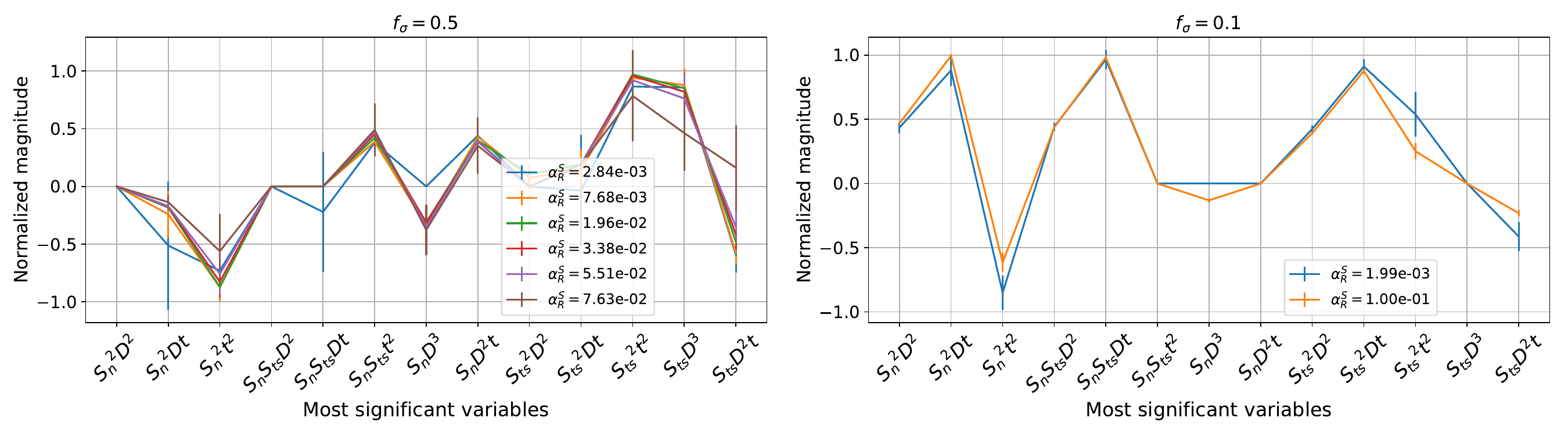}
    \caption{Each panel shows, for a fixed $f_\sigma$, the terms with largest magnitude in the matrix $W_\mathrm{out}$ of the RCs. The $10$ dominant terms of $W_\mathrm{out}$ are extracted from each of the $30$ RCs. Their magnitude is then normalized to that of the largest term and only terms that are most significant in more than $20$ out of $30$ RCs are retained. Each curve corresponds to a value of the parameter $\overline{E_a}$ (so that $\alpha_R^S$ lies between $0.002$ and $0.1$) and they show the average magnitude of each term across the $30$ RCs and the errorbars show the corresponding standard deviations.}
    \label{fig:interpretation_S}
\end{figure*}

Fig.~\ref{fig:interpretation_S} shows the same type of curves as Fig.~\ref{fig:interpretation}, obtained from the matrix $W_\mathrm{out}$ of $30$ RCs trained on an S-transition of the AMOC, for two values of $f_\sigma$: $0.5$ and $0.1$. Each curve corresponds to a grid cell in Fig.~\ref{fig:S_diagrams}. For each value of $f_\sigma$, we display all curves for which the probability estimates $\alpha_R^S$ lies between $0.002$ and $0.1$. This corresponds to only two curves in panel b), because the jumps in transition probabilities between consecutive samples of $\overline{E_a}$ is larger for $f_\sigma=0.1$ than for $f_\sigma=0.5$ (as can be observed in Fig.~\ref{fig:S_diagrams}a). For both values of $f_\sigma$, all curves are very consistent, showing the ability of the RCs to capture some structure of the phase space and to converge, although they are constantly independently retrained. For $f_\sigma=0.5$, the curve corresponding to the largest $\alpha_R^S$ deviates from the others (on the right of panel a)), although it has large error bars. In both panels, the curves corresponding to the lowest transition probabilities also have large error bars, showing the limits of transfer learning with a large difference between consecutive samples of $\overline{E_a}$. In this case, the analytical results for the committor that can be extracted from these curves are:
\begin{equation}
    \begin{aligned}
        f_\sigma=0.5:\ \Phi_R^S \propto\ &-c_1 t^2\left(\frac{3S_n^2}{4}-S_{ts}^2+\frac{S_nS_{ts}}{2}\right) \\
        &- c_2 \frac{D^3}{4}(S_n-3S_{ts}) \\
        &+ c_3 \frac{Dt}{2}(S_n-S_{ts})\left(D-\frac{S_n+S_{ts}}{2}\right)\\
        f_\sigma=0.1:\ \Phi_R^S \propto\ &d_1  D\left(t+\frac{D}{2}\right)((S_n+S_{ts})^2-S_nS_{ts}) \\
        - &d_2 \frac{t^2}{2}\left(\frac{3S_n^2}{2}-S_{ts}^2\right) \\
        - &d_3 S_{ts}D^2t
    \end{aligned}
\end{equation}
for again specific ${\cal O}(1)$ coefficients $c_i$ and $d_i$. These expressions are difficult to relate to $\Phi_S^S$ because the latter only relies on Euclidean distances in phase space rather than specific physical variables. 

\section{Discussion}
\label{sec:discussion}

This is to our knowledge the first study to couple TAMS to a machine learning technique that estimates the committor function on-the-fly. On the one hand, \citet{Lucente2022} had already coupled Adaptive Multilevel Splitting (an algorithm similar to TAMS) to another machine learning technique, the Analogue Markov Chain, which estimated the committor beforehand, thus creating an additional computational cost. On the other hand, direct committor function estimation using machine learning has already been performed on complex climate data \citep{Miloshevich2023b}, but only for short-term probabilistic forecast. 

We showed that the committor function can be learned at no additional cost than needed to compute transition probabilities. We compared our results to a combination of TAMS with a physics-informed score function and showed that transition probabilities and mean first-passage times were consistent. In the case of F-transitions, TAMS-$R$ also yields on average a similar variance in the results as TAMS-$S$, except for very low noise. Regarding S-transitions, both methods remain consistent for transition probabilities within $1000$ years down to $10^{-6}$. Furthermore, TAMS-$R$ consistently converges in fewer iterations than TAMS-$S$, suggesting that, in this case, RCs can learn a more efficient score function than $\Phi_S^S$. Finally, we showed that the analytical interpretation of the learned committor is consistent in the case of the F-transition with the physics-informed score function.

Concerning the AMOC, we confirmed that the probability of a collapse within a time frame of $100$ years for the F-transition and $1000$ years for the S-transition is not negligible for a large range of model parameters. Moreover, our method yields transition paths that are consistent with those obtained with a physics-informed score function. Finally, some of the terms of the estimated expression of the committor indicate that the RC captures the salt-advection feedback as key in the collapse process. 

The next step is to study AMOC tipping in higher-dimensional models like Earth Models of Intermediate Complexity (EMICs) or even Global Climate Models (GCMs). Another rare-event algorithm \citep{DelMoral2005,Tailleur2006,Lecomte2007} has already been applied to EMICs \citep{Ragone2018,Ragone2021,Wouters2023} and AMOC tipping \citep{Cini2024}. However, it focuses on the sampling of rare events rather than computing transition probabilities. Another interesting data-driven approach is the application of the edge-tracking algorithm to AMOC tipping \citep{Mehling2024}. It finds the critical state through which an AMOC tipping would occur, using a limited number of simulations, and is thus tractable in higher dimensions. This approach provides insight into the dynamics of the system but does not quantify the tipping probability. 

Extending our method to complex climate models raises the question of its scaling with the dimension of the model. For TAMS-$R$, this question is twofold: the feasibility of simulating enough trajectories and the scaling of RC itself. Regarding the first point, TAMS is not more computationally demanding than any other rare-event algorithm so their recent successes in EMICs \citep{Cini2024} are promising for TAMS. When it comes to GCMs \citep{vanWesten2023}, running TAMS does not seem feasible yet. 

However, the current implementation of the RC method poses a scaling problem. Due to the monomials used as basis functions, the size of the feature vector grows exponentially with the number of variables. We can only consider a very limited number of variables, which may be too simple to describe AMOC tipping in a GCM. \citet{Miloshevich2023} showed that Convolutional Neural Networks (CNN) are suitable for learning the committor from maps of climate variables. However, training CNNs still requires a large amount of data. More generally, this raises the problem of the choice of basis functions for the RC. The same issue arises for another committor estimation method, the Dynamical Galerkin Approximation \citep{Thiede2019,Finkel2021,Jacques-Dumas2023}. \citet{Strahan2023} recently used a Feedforward Neural Network to solve the Feynman-Kac formula for the committor. It only requires short trajectories such as those obtained with TAMS but may not solve the issue of the amount of data needed for training. 

The interpretability of the RC could also prove useful on the question of dimensionality. We showed in Sect.~\ref{sec:interpret} that in the feature vector of the RC, only a dozen of the terms play a significant role in the estimate. If these can be detected early enough in the learning process, dropping the others would speed up the computation. More generally, neural networks are often considered to be black boxes and difficult to interpret. Here, important variables and their relationship can be highlighted, which may serve as a useful starting point to focus observations or experiments. The efficiency of TAMS may rely on sufficient insight into the model's dynamics \citep{Lestang_2018}, but we show here that an efficient TAMS can also provide such insight.

Finally, the initialization and training process of the RC must be refined. In complex models, transition probabilities may only be computed for a single model setup, which makes transfer learning unusable. The RC could then be initialized from a linear score function that makes it easy to generate surrogate data. Another important issue is the memory of the RC. The current transfer learning process may retain too much information corresponding to very different system parameters since we did not "clean" its memory during the computations. The same issue may arise if the initialization of the RC is very far from the actual committor. A possible solution is to reinitialize the RC at certain milestones, once a certain number of transitions have been reached, thus using "better" data.

\acknowledgments
This project has received funding from the European Union’s Horizon 2020 research and innovation program under the Marie Sklodowska-Curie grant agreement no. 956170. R.M. van Westen and H.A. Dijkstra received funding from the European Research Council through the ERC-AdG project TAOC (project 101055096, PI: Dijkstra). 

\datastatement
The Python implementation of both models, all methods, and the code that produces the result plots can be found at the following address: https://doi.org/10.5281/zenodo.10532514 \citep{code}.

\newpage

\appendix[A] 
\appendixtitle{TAMS algorithm}
TAMS depends on few parameters: the number $N$ of ensemble members to simulate, the number $n_c$ of trajectories
to discard at each iteration and the maximum simulation time $T_\mathrm{max}$. 

\begin{enumerate}
	\item Initialize $k=0$ and $w_0=1$. 
	\item Simulate $N$ trajectories $\left(\mathbf{X}^{(1)},\dots,\mathbf{X}^{(N)}\right)$ starting in set $A$ until $T_{\mathrm{max}}$ or until they reach set $B$. Repeat steps $3-8$ until at least $N-n_c$ (or $N$ if $n_c=1$) trajectories have reached $B$.
	\item Compute the score function $\Phi(\mathbf{x})$ on every point of each trajectory.
	\item For trajectory $i$, the maximum value of the score function is called $Q_i$.  Let $I_k=\{i\in[1,N]\ |\ Q_i\in\{Q^*_{1},...,Q^*_{n_c}\}\}$, where $\{Q^*_j\}_{j\in[1,n_c]}$ are the $n_c$ lowest values of $Q_i$. Since the system is discrete, $I$ may contain more than $n_c$ elements. 
    \item Set $k=k+1$ then $w_k = \left(1-\frac{\#I_k}{N}\right)w_{k-1}$. For every $i\in I_k$, repeat steps $7-9$.
	\item Randomly select a trajectory $\mathbf{X}^{(r)}$ (such that $r\notin I_k$). We call $\tau$ the first time so that $\Phi(\mathbf{X}^{(r)}(\tau))\geq Q_i$. 
    \item Set $\mathbf{X}^{(i)}([1,\tau]) = \mathbf{X}^{(r)}([1,\tau])$.
	\item Simulate the rest of $\mathbf{X}^{(i)}$ starting from $\mathbf{X}^{(r)}(\tau)$ until time $T_{\mathrm{max}}-\tau$ or until reaching $B$.
\end{enumerate}

Let $N_B$ the number of trajectories that reached $B$ and $K$ and the number of iterations. The estimated transition probability from $A$ to $B$ before $T_\mathrm{max}$ is given by 
\citep{Rolland2015,Rolland2022}:
\begin{equation}
    \hat{\alpha} = \frac{N_B}{N}w_K = \frac{N_B}{N}\prod_{k=1}^K w_k = \frac{N_B}{N}\prod_{k=0}^{K-1} \left(1-\frac{\#I_k}{N}\right)
\end{equation}

The optimal variance, obtained when the score function is the committor function reads:
\begin{equation}
\label{eq:id_var}
    \sigma^2_\mathrm{id}=\frac{<\hat{\alpha}>^2}{N}\left(<K>\frac{n_c}{N-n_c} + \frac{N-<N_B>}{<N_B>}\right)
\end{equation}
where $<\cdot>$ represents an average over $M$ runs of TAMS. 

\newpage

\appendix[B] 
\appendixtitle{Next-Generation Reservoir Computing}

Reservoir Computing \citep{Jaeger2001,Lukosevicius2009} is a kind of neural network designed to represent dynamical systems and temporal processes
better than feedforward networks. The idea is to transform a nonlinear problem in the original phase space into a linear problem by embedding input data into a larger-dimensional space where linear regression is performed. Classical reservoir computers are universal approximators \citep{Gonon2020} but pose certain problems. For instance, their performance can vary depending on the realizations of the reservoir, which is usually a fixed random matrix.

The ``Next-Generation'' Reservoir Computing (here abbreviated RC) developed by \citet{Gauthier2021} is also an universal approximator \citep{Gonon2020, Hart2021} but embeds input data into a large-dimensional space using a family of nonlinear basis functions instead of a random matrix. We provide below only the important steps to build an RC for our problem. Details of the architecture can be found in \citet{Gauthier2021} and in \citet{Jacques-Dumas2023} for the setup and choice of parameters.

The RC receives at each time step a vector $\mathbf{x}\in\mathbb{R}^4$ containing $S_n$, $S_{ts}$, $D$ and the time $t$. The family of non-linear functions consists of all products of power $p=4$ of these variables. The resulting feature vector $\mathbf{\tilde{x}}\in\mathbb{R}^d$, with $d=40$ is much smaller than in classical reservoir computing (where its dimension would be approximately $1,000$), making the training of RC more efficient. 

The output matrix $\mathbf{W}_\mathrm{out}\in\mathbb{R}^{2\times d}$ maps $\mathbf{\tilde{x}}$ to the target output $\mathbf{y}\in\mathbb{R}^2$. Since the committor is inaccessible, its values cannot be used as target for the linear regression. We only know whether $\mathbf{x}$ first leads the trajectory to $A$ or to $B$ or if $T_{\mathrm{max}}$ is reached before any of these sets. The target output is $\mathbf{y}=(y_1,y_2)$ where $y_1=1$ if $A$ or $T_{\mathrm{max}}$ is reached first and $0$ otherwise ; $y_2=1$ if $B$ is reached first and $0$ otherwise. A Softmax operation then transforms these predictions into probabilities, namely: ("probability to reach $A$ or $T_{\mathrm{max}}$ first", "probability to reach $B$ first"). The second is an estimate of the committor.  

Since $\tilde{\mathbf{x}}$ and $\mathbf{y}$ are matched through linear regression, an analytical expression of $\mathbf{W}_\mathrm{out}$ can be derived \citep{Lukosevicius2009,Gauthier2021}, only depending on the feature vectors $\mathbf{\tilde{X}}$ and associated labels $\mathbf{Y}$:
\begin{equation}
\label{eq:rc_eq_2}
\mathbf{W}_\mathrm{out} = \mathbf{Y} \mathbf{\tilde{X}}^\mathrm{T}(\mathbf{\tilde{X}}\mathbf{\tilde{X}}^\mathrm{T}+\gamma\mathbf{I})^{-1}
\end{equation}
where $\mathbf{I}$ is the identity matrix and $\gamma$ is the Tikhonov regularization parameter. 

Finally, an important advantage of using RC is their ability to be trained online, for instance to update $W_\mathrm{out}$ at each iteration of TAMS.
Let $\mathbf{\tilde{X}}_k$ and $\mathbf{Y}_k$ be all the feature vectors and all the labels used up to iteration $k$. If new data become available, the newly computed $\mathbf{\tilde{x}}$ and $\mathbf{y}$ become new columns of $\mathbf{\tilde{X}}_k$ and $\mathbf{Y}_k$. The matrices $\mathbf{Y}_k\mathbf{\tilde{X}}_k^\mathrm{T}$ and $\mathbf{\tilde{X}}_k\mathbf{\tilde{X}}_k^\mathrm{T}$ are then updated:
\begin{equation}
\begin{aligned}
	\mathbf{Y}_{k+1}\mathbf{\tilde{X}}_{k+1}^\mathrm{T} &= \mathbf{Y}_k\mathbf{\tilde{X}}_k^\mathrm{T} + \mathbf{y}\mathbf{\tilde{x}}^\mathrm{T}\\
	\mathbf{\tilde{X}}_{k+1}\mathbf{\tilde{X}}_{k+1}^\mathrm{T} &= \mathbf{\tilde{X}}_k\mathbf{\tilde{X}}_k^\mathrm{T} + \mathbf{x}\mathbf{\tilde{x}}^\mathrm{T}
\end{aligned}
\end{equation}

\bibliographystyle{ametsocV6}
\bibliography{biblio}

\begin{thebibliography}{49}
\providecommand{\natexlab}[1]{#1}
\providecommand{\url}[1]{\texttt{#1}}
\renewcommand{\UrlFont}{\rmfamily}
\providecommand{\urlprefix}{URL }
\expandafter\ifx\csname urlstyle\endcsname\relax
  \providecommand{\doi}[1]{https://doi.org/\discretionary{}{}{}#1}\else
  \providecommand{\doi}{https://doi.org/\discretionary{}{}{}\begingroup \urlstyle{rm}\Url}\fi
\providecommand{\eprint}[2][]{\url{#2}}

\bibitem[{Armstrong~McKay et~al.(2022)}]{McKay2022}
Armstrong~McKay, D.~I., and Coauthors, 2022: {Exceeding 1.5$^{\circ}$C global warming could trigger multiple climate tipping points}. \textit{Science}, \textbf{377~(6611)}, eabn7950.

\bibitem[{Baars et~al.(2021)Baars, Castellana, Wubs,, and Dijkstra}]{Baars2021}
Baars, S., D.~Castellana, F.~Wubs, and H.~Dijkstra, 2021: Application of adaptive multilevel splitting to high-dimensional dynamical systems. \textit{Journal of Computational Physics}, \textbf{424}, 109\,876, \doi{https://doi.org/10.1016/j.jcp.2020.109876}.

\bibitem[{Bryden et~al.(2011)Bryden, King,, and McCarthy}]{Bryden2011}
Bryden, H.~L., B.~A. King, and G.~D. McCarthy, 2011: South atlantic overturning circulation at 24?s. \textit{Journal of Marine Research}, \textbf{69~(1)}, 38--56.

\bibitem[{{Castellana} et~al.(2019){Castellana}, {Baars}, {Wubs},, and {Dijkstra}}]{Castellana}
{Castellana}, D., S.~{Baars}, F.~W. {Wubs}, and H.~A. {Dijkstra}, 2019: {Transition Probabilities of Noise-induced Transitions of the Atlantic Ocean Circulation}. \textit{Scientific Reports}, \textbf{9}, 20284, \doi{10.1038/s41598-019-56435-6}.

\bibitem[{C\'{e}rou et~al.(2019)C\'{e}rou, Delyon, Guyader,, and Rousset}]{Cerou2019}
C\'{e}rou, F., B.~Delyon, A.~Guyader, and M.~Rousset, 2019: On the asymptotic normality of adaptive multilevel splitting. \textit{SIAM/ASA Journal on Uncertainty Quantification}, \textbf{7~(1)}, 1--30, \doi{10.1137/18M1187477}, \eprint{https://doi.org/10.1137/18M1187477}.

\bibitem[{Chang et~al.(2008)}]{Chang2008}
Chang, P., and Coauthors, 2008: Oceanic link between abrupt changes in the north atlantic ocean and the african monsoon. \textit{Nature Geoscience}, \textbf{1}, 444--448.

\bibitem[{Cimatoribus et~al.(2014)Cimatoribus, Drijfhout,, and Dijkstra}]{Cimatoribus}
Cimatoribus, A.~A., S.~S. Drijfhout, and H.~A. Dijkstra, 2014: Meridional overturning circulation: stability and ocean feedbacks in a box model. \textit{Climate Dynamics}, \textbf{42~(1)}, 311--328, \doi{10.1007/s00382-012-1576-9}.

\bibitem[{Cini et~al.(2024)Cini, Zappa, Ragone,, and Corti}]{Cini2024}
Cini, M., G.~Zappa, F.~Ragone, and S.~Corti, 2024: Simulating amoc tipping driven by internal climate variability with a rare event algorithm. \textit{npj Climate and Atmospheric Science}, \textbf{7}, \doi{10.1038/s41612-024-00568-7}.

\bibitem[{Collins et~al.(2013)}]{Collins2013}
Collins, M., and Coauthors, 2013: Long-term climate change: Projections, commitments and irreversibility. \urlprefix\url{https://api.semanticscholar.org/CorpusID:129640269}.

\bibitem[{Dijkstra(2023)}]{Dijkstra2023}
Dijkstra, H.~A., 2023: The role of conceptual models in climate research. \textit{Physica D: Nonlinear Phenomena}, 133984.

\bibitem[{Finkel et~al.(2021)Finkel, Webber, Gerber, Abbot,, and Weare}]{Finkel2021}
Finkel, J., R.~J. Webber, E.~P. Gerber, D.~S. Abbot, and J.~Weare, 2021: Learning forecasts of rare stratospheric transitions from short simulations. \textit{Monthly Weather Review}, \textbf{149~(11)}, 3647 -- 3669, \doi{10.1175/MWR-D-21-0024.1}.

\bibitem[{Freidlin and Wentzell(1998)Freidlin, and Wentzell}]{Freidlin1998}
Freidlin, M.~I., and A.~D. Wentzell, 1998: \textit{Random Perturbations}, 15--43. Springer New York, New York, NY, \doi{10.1007/978-1-4612-0611-8_2}, \urlprefix\url{https://doi.org/10.1007/978-1-4612-0611-8_2}.

\bibitem[{Garzoli et~al.(2013)Garzoli, Baringer, Dong, Perez,, and Yao}]{Garzoli2013}
Garzoli, S., M.~Baringer, S.~Dong, R.~Perez, and Q.~Yao, 2013: South atlantic meridional fluxes. \textit{Deep Sea Research Part I: Oceanographic Research Papers}, \textbf{71}, 21–32, \doi{10.1016/j.dsr.2012.09.003}.

\bibitem[{Gauthier et~al.(2021)Gauthier, Bollt, Griffith,, and Barbosa}]{Gauthier2021}
Gauthier, D.~J., E.~Bollt, A.~Griffith, and W.~A.~S. Barbosa, 2021: Next generation reservoir computing. \textit{Nature Communications}, \textbf{12~(1)}, 5564, \doi{10.1038/s41467-021-25801-2}.

\bibitem[{Gonon and Ortega(2020)Gonon, and Ortega}]{Gonon2020}
Gonon, L., and J.-P. Ortega, 2020: Reservoir computing universality with stochastic inputs. \textit{IEEE Transactions on Neural Networks and Learning Systems}, \textbf{31~(1)}, 100--112, \doi{10.1109/TNNLS.2019.2899649}.

\bibitem[{Hall and Ronald(2001)Hall, and Ronald}]{Hall2001}
Hall, A., and S.~Ronald, 2001: An abrupt climate event in a coupled ocean-atmosphere simulation without external forcing. \textit{Nature}, \textbf{409}, 171--4, \doi{10.1038/35051544}.

\bibitem[{Hart et~al.(2021)Hart, Hook,, and Dawes}]{Hart2021}
Hart, A.~G., J.~L. Hook, and J.~H. Dawes, 2021: Echo state networks trained by tikhonov least squares are l2($\mu$) approximators of ergodic dynamical systems. \textit{Physica D: Nonlinear Phenomena}, \textbf{421}, 132\,882, \doi{https://doi.org/10.1016/j.physd.2021.132882}.

\bibitem[{Jackson et~al.(2015)Jackson, Kahana, Graham, Ringer, Woollings, Mecking,, and Wood}]{Jackson2015}
Jackson, L., R.~Kahana, T.~Graham, M.~Ringer, T.~Woollings, J.~Mecking, and R.~Wood, 2015: Global and european climate impacts of a slowdown of the amoc in a high resolution gcm. \textit{Climate Dynamics}, \textbf{45}, 1--18, \doi{10.1007/s00382-015-2540-2}.

\bibitem[{Jacques-Dumas(2024)}]{code}
Jacques-Dumas, V., 2024: Estimation of amoc transition probabilities using a machine learning based rare-event algorithm [code]. Zenodo, \doi{10.5281/zenodo.10532514}.

\bibitem[{Jacques-Dumas et~al.(2023)Jacques-Dumas, van Westen, Bouchet,, and Dijkstra}]{Jacques-Dumas2023}
Jacques-Dumas, V., R.~M. van Westen, F.~Bouchet, and H.~A. Dijkstra, 2023: Data-driven methods to estimate the committor function in conceptual ocean models. \textit{Nonlinear Processes in Geophysics}, \textbf{30~(2)}, 195--216, \doi{10.5194/npg-30-195-2023}.

\bibitem[{Jaeger(2001)}]{Jaeger2001}
Jaeger, H., 2001: The" echo state" approach to analysing and training recurrent neural networks-with an erratum note'. \textit{Bonn, Germany: German National Research Center for Information Technology GMD Technical Report}, \textbf{148}.

\bibitem[{Kriegler et~al.(2009)Kriegler, Hall, Held, Dawson,, and Schellnhuber}]{Kriegler2009}
Kriegler, E., J.~W. Hall, H.~Held, R.~Dawson, and H.~J. Schellnhuber, 2009: Imprecise probability assessment of tipping points in the climate system. \textit{Proceedings of the National Academy of Sciences}, \textbf{106~(13)}, 5041--5046, \doi{10.1073/pnas.0809117106}, \eprint{https://www.pnas.org/doi/pdf/10.1073/pnas.0809117106}.

\bibitem[{Lecomte and Tailleur(2007)Lecomte, and Tailleur}]{Lecomte2007}
Lecomte, V., and J.~Tailleur, 2007: A numerical approach to large deviations in continuous time. \textit{Journal of Statistical Mechanics: Theory and Experiment}, \textbf{2007~(03)}, P03\,004, \doi{10.1088/1742-5468/2007/03/P03004}.

\bibitem[{Lenton et~al.(2008)Lenton, Held, Kriegler, Hall, Lucht, Rahmstorf,, and Schellnhuber}]{Lenton2008}
Lenton, T.~M., H.~Held, E.~Kriegler, J.~W. Hall, W.~Lucht, S.~Rahmstorf, and H.~J. Schellnhuber, 2008: Tipping elements in the earth's climate system. \textit{Proceedings of the National Academy of Sciences}, \textbf{105~(6)}, 1786--1793, \doi{10.1073/pnas.0705414105}, \eprint{https://www.pnas.org/doi/pdf/10.1073/pnas.0705414105}.

\bibitem[{Lestang et~al.(2018)Lestang, Ragone, Br{\'{e}}hier, Herbert,, and Bouchet}]{Lestang_2018}
Lestang, T., F.~Ragone, C.-E. Br{\'{e}}hier, C.~Herbert, and F.~Bouchet, 2018: Computing return times or return periods with rare event algorithms. \textit{Journal of Statistical Mechanics: Theory and Experiment}, \textbf{2018~(4)}, 043\,213, \doi{10.1088/1742-5468/aab856}.

\bibitem[{Li and Born(2019)Li, and Born}]{Li2019}
Li, C., and A.~Born, 2019: Coupled atmosphere-ice-ocean dynamics in dansgaard-oeschger events. \textit{Quaternary Science Reviews}, \textbf{203}, 1--20, \doi{10.1016/j.quascirev.2018.10.031}.

\bibitem[{Lucente et~al.(2022)Lucente, Rolland, Herbert,, and Bouchet}]{Lucente2022}
Lucente, D., J.~Rolland, C.~Herbert, and F.~Bouchet, 2022: Coupling rare event algorithms with data-based learned committor functions using the analogue markov chain. \textit{Journal of Statistical Mechanics: Theory and Experiment}, \textbf{2022~(8)}, 083\,201, \doi{10.1088/1742-5468/ac7aa7}.

\bibitem[{Lukoševičius and Jaeger(2009)Lukoševičius, and Jaeger}]{Lukosevicius2009}
Lukoševičius, M., and H.~Jaeger, 2009: Reservoir computing approaches to recurrent neural network training. \textit{Computer Science Review}, \textbf{3~(3)}, 127--149, \doi{https://doi.org/10.1016/j.cosrev.2009.03.005}.

\bibitem[{Mehling et~al.(2024)Mehling, Börner,, and Lucarini}]{Mehling2024}
Mehling, O., R.~Börner, and V.~Lucarini, 2024: Limits to predictability of the asymptotic state of the atlantic meridional overturning circulation in a conceptual climate model. \textit{Physica D: Nonlinear Phenomena}, \textbf{459}, 134\,043, \doi{https://doi.org/10.1016/j.physd.2023.134043}.

\bibitem[{Miloshevich et~al.(2023{\natexlab{a}})Miloshevich, Cozian, Abry, Borgnat,, and Bouchet}]{Miloshevich2023}
Miloshevich, G., B.~Cozian, P.~Abry, P.~Borgnat, and F.~Bouchet, 2023{\natexlab{a}}: Probabilistic forecasts of extreme heatwaves using convolutional neural networks in a regime of lack of data. \textit{Phys. Rev. Fluids}, \textbf{8}, 040\,501, \doi{10.1103/PhysRevFluids.8.040501}.

\bibitem[{Miloshevich et~al.(2023{\natexlab{b}})Miloshevich, Lucente, Yiou,, and Bouchet}]{Miloshevich2023b}
Miloshevich, G., D.~Lucente, P.~Yiou, and F.~Bouchet, 2023{\natexlab{b}}: Extreme heatwave sampling and prediction with analog markov chain and comparisons with deep learning. \eprint{2307.09060}.

\bibitem[{Moral and Garnier(2005)Moral, and Garnier}]{DelMoral2005}
Moral, P.~D., and J.~Garnier, 2005: Genealogical particle analysis of rare events. \textit{The Annals of Applied Probability}, \textbf{15~(4)}, 2496--2534.

\bibitem[{Parsons et~al.(2014)Parsons, Yin, Overpeck, Stouffer,, and Malyshev}]{Parsons2014}
Parsons, L.~A., J.~Yin, J.~T. Overpeck, R.~J. Stouffer, and S.~Malyshev, 2014: Influence of the atlantic meridional overturning circulation on the monsoon rainfall and carbon balance of the american tropics. \textit{Geophysical Research Letters}, \textbf{41~(1)}, 146--151, \doi{https://doi.org/10.1002/2013GL058454}, \eprint{https://agupubs.onlinelibrary.wiley.com/doi/pdf/10.1002/2013GL058454}.

\bibitem[{Ragone and Bouchet(2021)Ragone, and Bouchet}]{Ragone2021}
Ragone, F., and F.~Bouchet, 2021: Rare event algorithm study of extreme warm summers and heatwaves over europe. \textit{Geophysical Research Letters}, \textbf{48~(12)}, e2020GL091\,197, \doi{https://doi.org/10.1029/2020GL091197}, \eprint{https://agupubs.onlinelibrary.wiley.com/doi/pdf/10.1029/2020GL091197}.

\bibitem[{Ragone et~al.(2018)Ragone, Wouters,, and Bouchet}]{Ragone2018}
Ragone, F., J.~Wouters, and F.~Bouchet, 2018: Computation of extreme heat waves in climate models using a large deviation algorithm. \textit{Proceedings of the National Academy of Sciences}, \textbf{115~(1)}, 24--29, \doi{10.1073/pnas.1712645115}, \eprint{https://www.pnas.org/doi/pdf/10.1073/pnas.1712645115}.

\bibitem[{Rahmstorf(1996)}]{Rahmstorf1996}
Rahmstorf, S., 1996: On the freshwater forcing and transport of the atlantic thermohaline circulation. \textit{Climate Dynamics}, \textbf{12~(12)}, 799--811, \doi{10.1007/s003820050144}.

\bibitem[{Rolland(2022)}]{Rolland2022}
Rolland, J., 2022: Collapse of transitional wall turbulence captured using a rare events algorithm. \textit{Journal of Fluid Mechanics}, \textbf{931}, A22, \doi{10.1017/jfm.2021.957}.

\bibitem[{Rolland and Simonnet(2015)Rolland, and Simonnet}]{Rolland2015}
Rolland, J., and E.~Simonnet, 2015: Statistical behaviour of adaptive multilevel splitting algorithms in simple models. \textit{Journal of Computational Physics}, \textbf{283}, 541–558, \doi{10.1016/j.jcp.2014.12.009}.

\bibitem[{Roquet and Wunsch(2022)Roquet, and Wunsch}]{Roquet2022}
Roquet, F., and C.~Wunsch, 2022: The atlantic meridional overturning circulation and its hypothetical collapse. \textit{Tellus A: Dynamic Meteorology and Oceanography}, \textbf{74}, 393--398, \doi{10.16993/tellusa.679}.

\bibitem[{Soons et~al.(2023)Soons, Grafke,, and Dijkstra}]{Soons2023}
Soons, J., T.~Grafke, and H.~A. Dijkstra, 2023: Optimal transition paths for amoc collapse and recovery in a stochastic box model. \eprint{2311.12734}.

\bibitem[{Stommel(1961)}]{Stommel1961}
Stommel, H., 1961: Thermohaline convection with two stable regimes of flow. \textit{Tellus}, \textbf{13~(2)}, 224--230, \doi{https://doi.org/10.1111/j.2153-3490.1961.tb00079.x}, \eprint{https://onlinelibrary.wiley.com/doi/pdf/10.1111/j.2153-3490.1961.tb00079.x}.

\bibitem[{Strahan et~al.(2023)Strahan, Finkel, Dinner,, and Weare}]{Strahan2023}
Strahan, J., J.~Finkel, A.~R. Dinner, and J.~Weare, 2023: Predicting rare events using neural networks and short-trajectory data. \textit{Journal of computational physics}, \textbf{488}, 112\,152, \doi{https://doi.org/10.1016/j.jcp.2023.112152}.

\bibitem[{Tailleur and Kurchan(2006)Tailleur, and Kurchan}]{Tailleur2006}
Tailleur, J., and J.~Kurchan, 2006: Probing rare physical trajectories with lyapunov weighted dynamics. \textit{Nature Physics}, \textbf{3}, 203, \doi{10.1038/nphys515}.

\bibitem[{Thiede et~al.(2019)Thiede, Giannakis, Dinner,, and Weare}]{Thiede2019}
Thiede, E.~H., D.~Giannakis, A.~R. Dinner, and J.~Weare, 2019: {Galerkin approximation of dynamical quantities using trajectory data}. \textit{The Journal of Chemical Physics}, \textbf{150~(24)}, 244\,111, \doi{10.1063/1.5063730}, \eprint{https://pubs.aip.org/aip/jcp/article-pdf/doi/10.1063/1.5063730/15560137/244111\_1\_online.pdf}.

\bibitem[{van Westen and Dijkstra(2023)van Westen, and Dijkstra}]{vanWesten2023}
van Westen, R.~M., and H.~A. Dijkstra, 2023: Asymmetry of amoc hysteresis in a state-of-the-art global climate model. \textit{Geophysical Research Letters}, \textbf{50~(22)}, e2023GL106\,088, \doi{https://doi.org/10.1029/2023GL106088}, \eprint{https://agupubs.onlinelibrary.wiley.com/doi/pdf/10.1029/2023GL106088}.

\bibitem[{van Westen et~al.(2024)van Westen, Kliphuis,, and Dijkstra}]{vanWesten2024}
van Westen, R.~M., M.~Kliphuis, and H.~A. Dijkstra, 2024: Physics-based early warning signal shows that amoc is on tipping course. \textit{Science Advances}, \textbf{10~(6)}, eadk1189, \doi{10.1126/sciadv.adk1189}, \eprint{https://www.science.org/doi/pdf/10.1126/sciadv.adk1189}.

\bibitem[{Vettoretti et~al.(2022)Vettoretti, Ditlevsen, Jochum,, and Rasmussen}]{Vettoretti2022}
Vettoretti, G., P.~Ditlevsen, M.~Jochum, and S.~Rasmussen, 2022: Atmospheric co2 control of spontaneous millennial-scale ice age climate oscillations. \textit{Nature Geoscience}, \textbf{15}, 1--7, \doi{10.1038/s41561-022-00920-7}.

\bibitem[{Weijer et~al.(2019)}]{Weijer2019}
Weijer, W., and Coauthors, 2019: {Stability of the Atlantic Meridional Overturning Circulation: A review and synthesis}. \textit{Journal of Geophysical Research: Oceans}, \textbf{124~(8)}, 5336--5375.

\bibitem[{Wouters et~al.(2023)Wouters, Schiemann,, and Shaffrey}]{Wouters2023}
Wouters, J., R.~K.~H. Schiemann, and L.~C. Shaffrey, 2023: Rare event simulation of extreme european winter rainfall in an intermediate complexity climate model. \textit{Journal of Advances in Modeling Earth Systems}, \textbf{15~(4)}, e2022MS003\,537, \doi{https://doi.org/10.1029/2022MS003537}, \eprint{https://agupubs.onlinelibrary.wiley.com/doi/pdf/10.1029/2022MS003537}.

\end{thebibliography}

\end{document}